\documentclass[twoside,11pt]{article}

\usepackage[preprint]{jmlr2e}

\usepackage{blindtext}
\usepackage{algorithm}
\usepackage{algpseudocode}

\usepackage{lastpage}
\jmlrheading{26}{2025}{1-\pageref{LastPage}}{9/24; Revised
11/25}{12/25}{24-1584}{John C. Yannotty, Thomas J. Santner, Bo Li, and Matthew T. Pratola}

\ShortHeadings{Combining Climate Models using Bayesian Regression Trees
and Random Paths}{Yannotty, Santner, Li, and Pratola}
\firstpageno{1}

\usepackage{indentfirst}
\usepackage{amsmath,amsfonts,amssymb,mathtools}
\usepackage{comment}
\usepackage{arydshln} 
\usepackage{setspace}

\newtheorem{theorem_label}{Theorem}[section]

\newtheorem{definition_label}[theorem]{Definition}

\newcommand{\ith}{i\text{th}}
\newcommand{\jth}{j\text{th}}

\newcommand{\lth}{l\text{th}}

\newcommand{\vth}{v\text{th}}
\newcommand{\bth}{b\text{th}}

\newcommand{\dth}{d\text{th}}

\newcommand{\Cset}{\mathcal{C}}
\newcommand{\Iset}{\mathcal{I}}

\newcommand{\X}{\mathcal{X}}

\newcommand{\R}{\mathbb{R}}

\newcommand{\muvec}{\boldsymbol\mu}
\newcommand{\wvec}{\boldsymbol w}
\newcommand{\hvec}{\boldsymbol h}
\newcommand{\psivec}{\boldsymbol \psi}
\newcommand{\uvec}{\boldsymbol u}

\newcommand{\zvec}{\boldsymbol z}
\newcommand{\yvec}{\boldsymbol y}
\newcommand{\fhatvec}{\boldsymbol{\hat{f}}}
\newcommand{\gvec}{\boldsymbol g}

\newcommand{\xvec}{\boldsymbol x}
\newcommand{\rvec}{\boldsymbol r}

\newcommand{\onevec}{\boldsymbol 1}

\usepackage{natbib}

\setcitestyle{authoryear,open={(}, close = {)}} 
\usepackage{wrapfig}
\usepackage[section]{placeins}
\usepackage{url}

\usepackage{color}
\usepackage{soul} 

\begin{document}

\def\spacingset#1{\renewcommand{\baselinestretch}%
{#1}\small\normalsize} \spacingset{1}

\title{\bf Combining  Climate Models using Bayesian Regression Trees and Random Paths}

\author{\name John C. Yannotty \email yannotty@battelle.org \\
       \addr Battelle Memorial Institute\\
       Columbus, OH 43201, USA
       \AND
       \name Thomas J. Santner \email santner.1@osu.edu \\
       \addr Department of Statistics\\
       The Ohio State University\\
       Columbus, OH 43210, USA
       \AND
       \name Bo Li \email bol@wustl.edu  \\
       \addr Department of Statistics and Data Science\\
       Washington University in St. Louis \\
       St. Louis, MO 63130, USA
       \AND
       \name Matthew T. Pratola \email mpratola@iu.edu  \\
       \addr Department of Statistics\\
       Indiana University\\
       Bloomington, IN 47405, USA}
\editor{Chris Oats}
\maketitle
\bigskip
\begin{abstract}

General circulation models (GCMs) are essential tools for climate studies. Such climate models may have varying accuracy across the
input domain, but no model is uniformly best. One can
improve climate model prediction performance by integrating multiple models using input-dependent weights. Weight functions modeled using Bayesian Additive Regression Trees (BART) were recently shown to be useful in nuclear physics applications. However,
a restriction of that approach was the piecewise constant weight functions. To smoothly integrate multiple climate models, we propose a new tree-based model, Random Path BART (RPBART), that incorporates random path assignments in BART to produce smooth weight functions and smooth predictions, all in a matrix-free formulation. RPBART requires a more complex prior specification, for which we introduce a semivariogram to guide hyperparameter selection. This approach is easy to interpret, computationally cheap, and avoids expensive cross-validation.
Finally, we propose a posterior projection technique to enable detailed analysis of the
fitted weight functions. This allows us to identify a sparse set of climate
models that recovers the underlying system within a given spatial region as well as quantifying model discrepancy given the available model set. Our method is demonstrated on an ensemble of 8 GCMs modeling average monthly surface temperature.

\end{abstract}

\noindent%
\vfill

\newpage

\section{Introduction}

Complex natural phenomena are often modeled using computer simulators -- i.e. models that incorporate theoretical knowledge to approximate an underlying system. In climate applications, these simulators, or General Circulation Models (GCMs), have been widely used to understand a variety of climate features such as temperature or precipitation \citep{eyring2016overview}. Many GCMs have been developed over time and each tends to have varying fidelity across different subregions of the world. Hence, no universally best model exists. 

Multi-model ensembles are often used to help improve global prediction of physical systems. A common approach is to explicitly combine the outputs from $K$ different GCMs using a linear combination or weighted average, usually in a pointwise manner for a specific latitude and longitude. For example, \citet{knutti2017climate} and \citet{tebaldi2007use} define performance-based weights. \cite{giorgi2002calculation} define Reliability Ensemble Averaging (REA), which weight the GCMs based on model performance and model convergence. 
Traditional regression-based models assume the underlying process can be modeled as a linear combination of the individual GCMs \citep{tebaldi2007use}.  \citet{vrac2024distribution} propose a density-based approach that combines the individual cumulative density functions from the $K$ models. 
Each of these methods estimate the weights in vastly different ways, however, they all derive location-specific weights based on the information at a given latitude and longitude location. 

Alternative approaches implicitly combine GCMs to estimate the mixed-prediction. For example, \cite{harris2023multimodel} model the underlying system as a Gaussian process with a deep neural network kernel, and then employ Gaussian process regression to combine multiple climate models.  \cite{sansom2021constraining} define a Bayesian hierarchical model that assumes a co-exchangeable relationship between climate models and the real world process.  


Global approaches such as Bayesian Model Averaging (BMA) \citep{bma_lr} and model stacking \citep{breiman_stacking, cs, clydec2013bayesian}, have not seen wide adoption in climate applications. These approaches use scalar weights to combine  simulator outputs. The corresponding weights in such schemes are meant to reflect the overall accuracy of each model, where larger weights indicate better performance. 

More recent advances consider localized weights, where the weights are explicitly modeled as functions over the input domain. The outputs from each model are then combined, or mixed, using weight functions that reflect each individual model's local predictive accuracy relative to the others in the model set. This approach is often referred to as model mixing and enables more effective learning of local information than pointwise methods without degenerating to  overly simplistic global weighting. One key challenge in model mixing is specifying the relationship between the inputs and the weight values. Some specific approaches for modeling the weights use linear basis functions \citep{sill2009feature}, generalized linear models \citep{hs}, neural networks \citep{COSCRATO2020141}, calibration-based weighting \citep{band2021}, Dirichlet-based weights \citep{kejzlar2023local}, precision weighting \citep{semposki2022uncertainties}, and Bayesian Additive Regression Trees (BART) \citep{yannotty2023model}. These approaches are conceptually related but differ in the assumed functional relationship for the weights and the capability to quantify uncertainties.

The BART approach for modeling the $K$-dimensional vector of weight functions is attractive due to its non-parametric formulation which avoids the need for  user-specified basis functions. Specifically, the BART approach defines a set of prior tree bases which are adaptively learned based on the information in the model set and the observational data. However, the resulting weight functions of BART are piecewise constant, resulting in the primary drawback of this approach: the weight functions and  predictions of the system are discontinuous. This is a noticeable limitation when smoothness is desirable. The univariate regression extension Soft BART (SBART) \citep{linero_sbart} allows smooth predictions using tree bases. However, SBART is better suited to modeling scalar responses rather than a vector-valued quantity. 
Additionally, \cite{linero_sbart} propose a set of default priors for the SBART model, which mitigates the need to select the values of hyperparameters and avoids a complex cross-validation study. These default settings may be insufficient in some applications and a more principled approach for prior calibration may be desired. Thus, directly applying the existing ``soft" regression tree methods to the current BART-based weight functions is both not straight-forward and computationally infeasible.

Our contributions to this research are as follows. First, we propose the Random Path BART (RPBART) model, which uses a latent variable approach to enable smooth predictions using an additive regression tree framework, all in a matrix-free formulation. Second, we introduce the random path model in the Bayesian Model Mixing framework (RPBART-BMM).
The RPBART-BMM model improves on the BART-based model mixing method introduced by \cite{yannotty2023model}, which was at times sensitive to  overfitting and thus provided poor uncertainty quantification in areas away from the training points. The proposed construction also introduces smoothness in a holistic way such that the induced smoothing is compatible with the localization effect of the learned tree structure.  
Additionally, we derive the prior semivariogram of the resulting model, allowing for principled yet efficient calibration of model prior hyperparameters with similar ease as the original BART proposal. We also introduce posterior projection methods for model mixing that can be used to better interpret the mixed-prediction and resulting weight functions in cases where all, some, or none of the models are locally useful for the system of interest. Finally, we demonstrate our methodology by combining the outputs from $K$ different GCMs to estimate the underlying mean process of the true system and gain insight as to where each GCM is locally accurate or inaccurate.  We demonstrate the enhanced performance of our method relative to competing methods in mixing GCMs.

The remainder of the paper is organized as follows. Section 2 reviews the relevant background literature relating to Bayesian regression trees. Section 3 outlines our novel Random Path (RPBART) methodology for scalar responses. Section 4 extends the RPBART model to Bayesian model mixing and introduces projections of the fitted weight functions. Section 5 applies our methods to GCMs, and Section 6 summarizes our contributions.

\section{Bayesian Regression Trees} 
\label{bart_background}
Bayesian regression trees were introduced in \cite{bcart, bart_2010,denison1998bayesian}. The most common ensemble approach is BART \citep{bart_2010}, which models the mean function using additive tree bases.   
 
\begin{figure}[t]
    \centering
    \includegraphics[width = 0.82\textwidth, height = 0.39\textwidth]{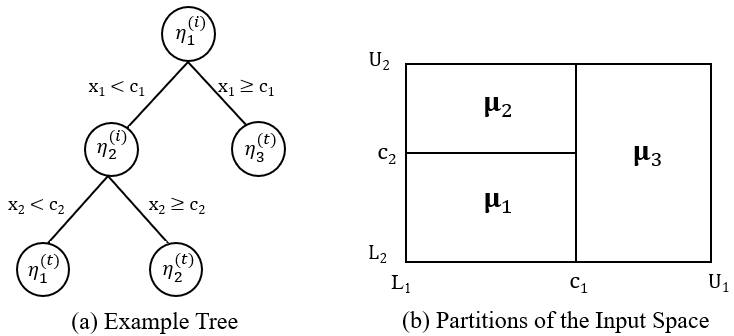}
    \caption{(Left) An example of a tree structure $T$ applied to a $2$-dimensional input space. The internal and terminal nodes of the tree are denoted with superscripts $(i)$ and $(t)$, respectively. Each internal node facilitates a binary split of the form $x_v < c_v$. (Right) The partitions of the rectangular input space $[L_1,U_1]\times[L_2,U_2]$, with associated terminal node parameters $\mu_{b}$, $b=1,2,3$. The function $g(\xvec;T,M)$ maps a given $\xvec$ to one of these three values.}
    \label{fig:bart_tree}
\end{figure}

A single tree $T$ consists of $B$ terminal nodes and $B-1$ internal nodes, which partitions a $p$-dimensional compact input space into $B$ disjoint subsets. Each internal node encodes a rule of the form $x_v < c_v$, where $v \in \{1,\ldots,p\},$ inducing a binary split along the $\vth$ dimension of the input space. Cutpoint $c_v$ is drawn from a discretized set over the interval $[L_v,U_v]$, the lower and upper bounds of the $\vth$ dimension of the input space, respectively. The terminal nodes are found in the bottom level of the tree. Terminal node $b$ corresponds to a unique partition of the input space and is assigned a unique terminal node parameter $\mu_b$. Therefore, a tree implicitly defines a function $g(\xvec;T,M)$ such that if $\xvec$ lies in the $\bth$ partition, then $g(\xvec;T,M) = \mu_b$, where $T$ denotes the tree topology and $M = \{\mu_1,\ldots,\mu_B\}$ denotes the set of terminal node parameters. By construction, $g(\xvec;T,M)$ is piecewise constant. Figure \ref{fig:bart_tree} displays an example with $B=3$ terminal nodes and a $2$-dimensional input space.
 
The tree topology prior, $\pi(T)$, defines the prior probability of node type (terminal or internal) and splitting rules for each internal node. The default prior is uninformative for the split rules and penalizes tree depth \citep{bart_2010} as $\pi(\eta \text{ is internal}) = \alpha(1+d_\eta)^{-\beta}$, where $\alpha$ and $\beta$ are tuning parameters and $d_\eta$ denotes the depth of $\eta$. The output of the tree corresponds to the set of the terminal node parameters $M$. Typically, a conditionally independent conjugate normal prior is assigned to each $\mu_1,\ldots,\mu_B$. These assumptions  simplify the  Markov Chain Monte Carlo (MCMC), particularly when working with a Gaussian likelihood.    
 
Samples from the posterior of $T,\ M$ and $\sigma^2$ are generated using MCMC. The tree topology is updated at each iteration by proposing a slight change to the existing tree structure \citep{bcart,pratola2016efficient}. The terminal node and variance parameters are then updated using Gibbs steps. 
 
The methodology outlined above is easily extended to an ensemble of $m$ trees, $T_1,\ldots,T_m,$ with terminal node parameter sets $M_1,\ldots,M_m$ \citep{bart_2010, hastie2000bayesian}. The mean function is then modeled as a sum-of-trees, $E[Y(\xvec) \mid \xvec] = \sum_{j = 1}^m g(\xvec; T_j, M_j),$ where $g(\xvec; T_j, M_j) = \sum_{b = 1}^{B_j} \mu_{bj} \; I(\xvec \in \X_{bj})$ and $I(\xvec \in \X_{bj})$ is an indicator function denoting the event that $\xvec$ maps to the $\bth$ partition of the input space by the $\jth$ tree. 
 
Generally, tree models result in a piecewise constant mean function. When continuity is desirable, soft regression trees \citep{irsoy2012soft,linero_sbart} are a useful alternative. A soft regression tree maps an observation with input $\xvec$ to a unique terminal node probabilistically given the tree topology and associated splitting rules. If $T_j$ has $B_j$ terminal nodes, the probability $\xvec$ maps to the $\bth$ terminal node is $\phi_{bj}(\xvec; T_j, \gamma_j)$ for $b=1,\ldots,B_j,$ for bandwidth parameter $\gamma_j$. The bandwidth $\gamma_j$ controls the amount of smoothing across terminal nodes. Assuming each input dimension is standardized to $x_v \in [0,1]$, SBART assigns an exponential prior to $\gamma_j$. Values of $\gamma_j\geq 0.1$ lead to a more global solution, while small values of $\gamma_j$ generate a localized fit similar to BART. The mean function of a soft regression tree is a weighted average of the terminal node parameters 
\begin{equation}
    g(\xvec; T_j, M_j) = \sum_{b = 1}^{B_j} \mu_{bj} \; \phi_{bj}(\xvec; T_j, \gamma_j). \label{sbart_sum_of_trees}
\end{equation}  
Traditional regression trees with deterministic paths correspond to $\phi_{bj}(\xvec; T_j,\gamma_j) = I(\xvec \in \X_{bj}).$
 
The terminal node parameters in (\ref{sbart_sum_of_trees}) need to be updated jointly during the MCMC. When these are scalars, this requires  inversion of a non-sparse $B_j\times B_j$ matrix \citep{linero_sbart}. When trees are regularized to maintain a shallow depth, this inversion is relatively inexpensive to compute. But when the parameters are $K$-dimensional, inverting the $KB_j \times KB_j$ matrix can become computationally expensive even with shallow trees. The proposed RPBART model alleviates these concerns.

\section{The Random Path Model}

We first propose our smooth, continuous, Random Path  model for standard BART (RPBART), which models a univariate mean function. Let $Y(\xvec_i)$ be an observable quantity from some unknown process  at input $\xvec_i$. Let $z_{bj}(\xvec_i)$ be the latent {\em random path indicator} for the event that the $\ith$ input is mapped to the $\bth$ terminal node in $T_j$.  Given the random path assignments, we assume $Y(\xvec_i)$ is modeled as 
\begin{align}
    Y(\xvec_i) \mid \{T_j,M_j,Z_j\}_{j=1}^m, \sigma^2 \;  &\sim \; N\Big(\sum_{j = 1}^m g(\xvec_i; T_j, M_j,Z_j), \sigma^2 \Big) \label{rpath_ydef},  \\
    g(\xvec_i; T_j, M_j, Z_j) &= \sum_{b=1}^{B_j} \mu_{bj}\; z_{bj}(\xvec_i). \label{rpath_gdef} 
\end{align}
Each observation is mapped to exactly one terminal node within $T_j$, thus $\sum_{b=1}^{B_j} z_{bj}(\xvec_i) = 1$ and $z_{bj}(\xvec_i) \in \{0,1\}$ for $b = 1,\ldots, B_j$ and $i=1,\ldots,n$. The set $Z_j$ is then defined as $Z_j = \{\zvec_j(\xvec_i) \}_{i = 1}^n$ where $\zvec_j(\xvec_i)$ is the $B_j$-dimensional vector of random path assignments for the $\ith$ input. 
Traditional BART with deterministic paths can be viewed as a special case of RPBART where $z_{bj}(\xvec_i) = I(\xvec_i \in \X_{bj})$. Conditional on $Z_j$, the output of the random path tree remains a piecewise constant form. However, taking the expectation of the sum-of-trees with respect to $Z_j$ results in a continuous function, similar to SBART (\ref{sbart_sum_of_trees}). 

\subsection{Prior Specification} \label{subsect:rpath_prior}
The original BART model depends on the tree structures, associated set of terminal node parameters, and error variance. We maintain the usual priors for each of these components,
\begin{align*}
    \mu_{bj} \mid T_j &\sim N(0, \tau^2), \quad T_j \sim \pi(T_j), \quad \sigma^2 \sim \nu\lambda/\chi^2_\nu,    
\end{align*}
where $b = 1,\ldots,B_j$ and $j=1,\ldots,m$. The values of $\tau$, $\lambda$, and $\nu$ can still be selected using similar methods as specified by \cite{bart_2010}.

RPBART introduces two new sets of parameters, $(Z_j, \gamma_j)$ for $j = 1,\ldots,m$.  For each $Z_j$, consider the path assignment for the $\ith$ observation, $\zvec_j(\xvec_i)$. Conditional on $T_j$, the $B_j$-dimensional random path vector for a given $\xvec_i$ is assigned a Multinomial prior 
\begin{equation}
    \zvec_j(\xvec_i) \mid T_j, \gamma_j \sim \text{Multinomial}\Big(1; \phi_{1j}(\xvec_i;T_j,\gamma_j),\ldots,\phi_{{B_j}j}(\xvec_i;T_j,\gamma_j) \Big) \label{zprior},
\end{equation}
where $\phi_{bj}(\xvec_i; T_j,\gamma_j)$ is the probability an observation with input $\xvec_i$ is mapped to the $\bth$ terminal node in $T_j$ or equivalently, the conditional probability that $z_{bj}(\xvec_i) = 1$. The bandwidth  parameter, $\gamma_j$, takes values within the interval $(0,1)$ and controls the degree of pooling across terminal nodes. As $\gamma_j$ increases, more information is shared across the terminal nodes, which leads to a less localized prediction. Since we confine $\gamma_j$ to the interval $(0,1)$ we assume $\gamma_j \sim \text{Beta}(\alpha_1,\alpha_2)$, for $j = 1,\ldots,m.$
This prior for $\gamma_j$ noticeably differs from the exponential prior in SBART \citep{linero_sbart}. The different modeling assumptions are guided by the design of the path probabilities, $\phi_{bj}(\xvec_i;T_j,\gamma_j)$, which are discussed in Section \ref{subsect:randpathprob}. Finally, we assume conditional independence given the set of $m$ trees which implies 
\begin{align}
    \pi\big(\sigma^2,\{T_j,M_j,Z_j,\gamma_j\}_{j=1}^m\big) &= \pi(\sigma^2)\prod_{j=1}^m \pi(M_j \mid T_j, Z_j)\;\pi(Z_j \mid T_j, \gamma_j)\;\pi(T_j)\pi(\gamma_j) \notag \\
    &=
    \pi(\sigma^2)\prod_{j=1}^m \pi(T_j)\;\pi(\gamma_j) \prod_{b = 1}^{B_j} \pi(\muvec_{bj} \mid T_j) \; \pi(Z_j \mid T_j, \gamma_j). \label{rpath_joint_prior}    
\end{align}
Furthermore, we assume mutual independence across the random path assignment vectors $\zvec_j(\xvec_i)$ apriori. Thus the set of vectors over the $n$ training points can be rewritten as
\begin{equation}
    \pi(Z_j \mid T_j, \gamma_j) = \prod_{i = 1}^n \prod_{b = 1}^{B_j} \Big(\phi_{bj}(\xvec_i; T_j,\gamma_j)\Big)^{z_{bj}(\xvec_i)}.
\end{equation}

\subsection{The Path Probabilities} \label{subsect:randpathprob}

The continuity in the mean prediction from each tree is driven by the $B_j$ path probabilities, $\phi_{bj}(\xvec_i; T_j, \gamma_j)$.  Recall, a tree model recursively partitions the input space into $B_j$ disjoint subregions using a sequence of splitting rules. We define the path from the root node, $\eta_{1j}^{(i)}$, to the $\bth$ terminal node, $\eta_{bj}^{(t)}$, in terms of the sequence of internal nodes that connect $\eta_{1j}^{(i)}$ and $\eta_{bj}^{(t)}$. To define $\phi_{bj}(\xvec_i; T_j, \gamma_j)$, we must consider the probability of visiting the internal nodes that form the path that connects $\eta_{1j}^{(i)}$ and $\eta_{bj}^{(t)}$.

Consider the tree with $B_j = 3$ terminal nodes (red, blue, green) and induced partition of the $1$-dimensional input space, $[-1,1]$, in Figure \ref{fig:rpath_phix_tree}. The path to  $\eta_{1j}^{(t)}$ (red) is defined by splitting left at the root node. Similarly, the path $\eta_{2j}^{(t)}$ (blue) is defined by splitting right at $\eta_{1j}^{(i)}$ and then left at $\eta_{2j}^{(i)}$. In the usual tree model, these splits happen deterministically. This means an observation with $x_1 < 0$ splits left and is thus mapped to $\eta_{1j}^{(t)}$ with probability 1. With random paths, an observation splits right with probability $\psi(\xvec; \cdot)$ and left with probability $1 - \psi(\xvec; \cdot)$. This adds another layer of stochasticity into the model.

\subsubsection{Defining the Splitting Probabilities}

Consider the $\dth$ internal node, $\eta^{(i)}_{dj}$, of $T_j$. Assume $\eta^{(i)}_{dj}$ splits on the rule $x_{v_{(dj)}} < c_{(dj)}$. Let $\psi(\xvec;v_{(dj)},c_{(dj)},\gamma_j)$ define the probability an observation with input $\xvec$ moves to the right child of $\eta^{(i)}_{dj}$. Further assume the cutpoint $c_{(dj)}$ is selected from the discretized subset $(L^{d}_v, U^{d}_v)\cap \Cset_v$, where $\Cset_v$ is the finite set of possible cutpoints for variable $v$, and $L^{d}_v$ and $U^{d}_v$ are the upper and lower bounds defined based on the previous splitting rules in the tree. The bounds, which are computed based on the information in $T_j$, are used to define a threshold which establishes a notion of ``closeness" between points. We incorporate this information into the definition of $\psi(\xvec;v_{(dj)},c_{(dj)},\gamma_j)$ by     
\begin{equation} \label{psi_rpath}
\psi(\xvec;v_{(dj)},c_{(dj)},\gamma_j)= 
    \begin{cases}
        1 - \frac{1}{2}\Big(1 - \frac{x_{v_{(dj)}} - c_{(dj)}}{\gamma_j(U^{d}_v - c_{(dj)})} \Big)^q_+ &  x_{v_{(dj)}} \geq c_{(dj)},  \\[7pt]
        \frac{1}{2}\Big(1 - \frac{c_{(dj)} - x_{v_{(dj)}}}{\gamma_j(c_{(dj)} - L^{d}_v)} \Big)^q_+ &  x_{v_{(dj)}} < c_{(dj)},
    \end{cases}
\end{equation}
where the expression $a_+ = \max\{a,0\}$ for any $a\in\R$ and $q$ is a shape parameter. This definition of $\psi(\xvec;v_{(dj)},c_{(dj)}, \gamma_j)$ restricts the probabilistic assignment to the left or right child of $\eta^{(i)}_{dj}$ to the interval $\Iset_{dj}(\gamma_j) := \big(c_{(dj)} - \gamma_j(c_{(dj)}-L^{d}_v), \; c_{(dj)} + \gamma_j(U^{d}_v-c_{(dj)}))$. Any observation with input $x_{v_{(dj)}}$ such that  $x_{v_{(dj)}} \in \Iset_{dj}(\gamma_j)$ has a non-zero chance of being assigned to either of the child nodes in the binary split. Meanwhile, $\psi(\xvec;v_{(dj)},c_{(dj)},\gamma_j)=0$ if $x_{v_{(dj)}}$ is less than the lower bound in $\Iset_{dj}(\gamma_j)$ or $\psi(\xvec;v_{(dj)},c_{(dj)},\gamma_j)=1$ if $x_{v_{(dj)}}$ is greater than the upper bound in $\Iset_{dj}(\gamma_j)$. This means the probabilistic assignment agrees with the deterministic split when $x_{v_{(dj)}} \notin \Iset_{dj}(\gamma_j)$. In other words, points farther away from the cutpoint deterministically move to the left or right, while points close to the cutpoint take a random move left or right.     

\begin{figure}[t]
    \centering
    \includegraphics[width = 1\textwidth, height = 0.4\textwidth]{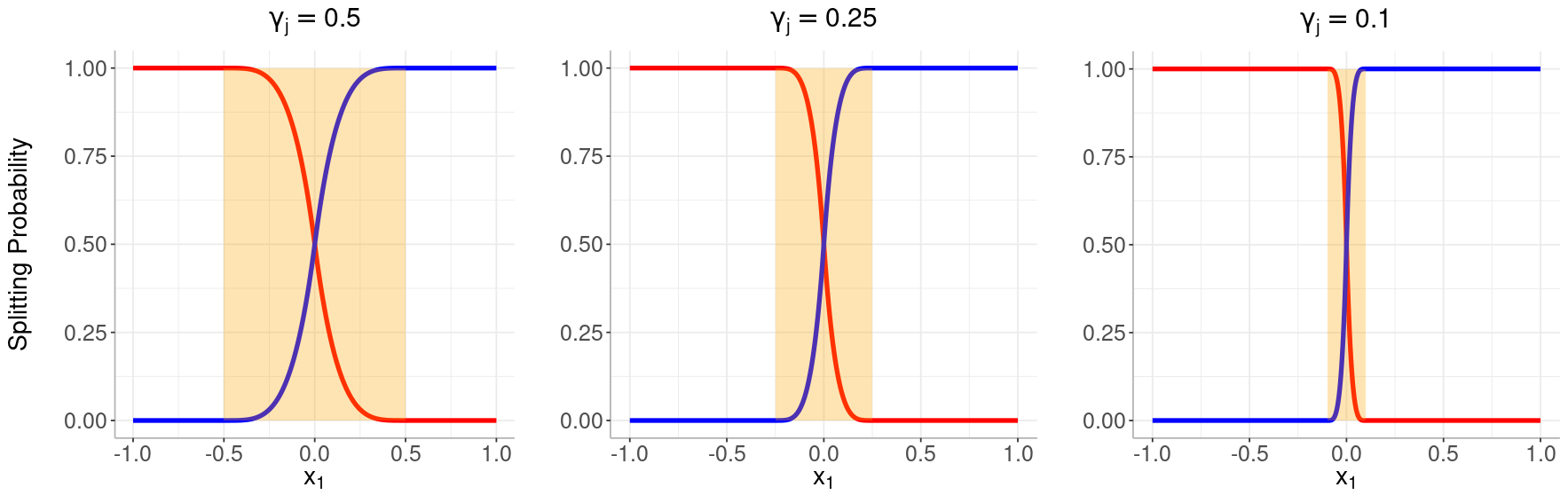}
    \caption{Three examples of $1-\psi(\xvec;1,0 , \gamma_j)$ (red) and $\psi(\xvec;1,0 , \gamma_j)$ (blue) with bandwidth parameters of $0.5$, $0.25$, and $0.1$, corresponding to the move that splits the interval $[L^{(1)}_1,U^{(1)}_1] = [-1,1]$ on the cutpoint $c_{1j} = 0$. The interval $\Iset_{1j}(\gamma_j)$ (orange) defines the interval where the random paths can disagree with the traditional deterministic paths.}
\label{fig:rpath_psix}
\end{figure}

To understand $\gamma_j$, consider the first split (at the root node) within a given a tree. Assume the root node splits the interval $[L_1^{(1)}, U_1^{(1)}] = [-1,1]$ using the rule $x_1 < 0$ (i.e. $v_{1j} = 1$ and $c_{1j} = 0$). Figure \ref{fig:rpath_psix} displays the probability of splitting left (red),  $1- \psi(\xvec;1,0,\gamma_j)$, and the probability of splitting right (blue), $\psi(\xvec;1,0,\gamma_j)$, for different values of $\gamma_j$ as a function of $x_1$. The orange region in each panel highlights the set $\Iset_{1j}(\gamma_j)$. The interval is wider for larger $\gamma_j$, which means a larger proportion of points can split left or right. Lower $\gamma_j$ results in steeper curves and
$\psi(\xvec;1,0,\gamma_j)$ starts to resemble the deterministic rule $I(x_{v_{(dj)}} \geq c_{(dj)})$.   

The splitting probabilities determine how to traverse the tree along the various paths created by the internal nodes. These probabilities set the foundation for computing the probability of reaching any of the $B_j$ terminal nodes within $T_j$.  
 
\subsubsection{Defining the Path Probabilities}

\begin{figure}[t]
    \centering
    \includegraphics[width = 0.95\textwidth, height = 0.375\textwidth]{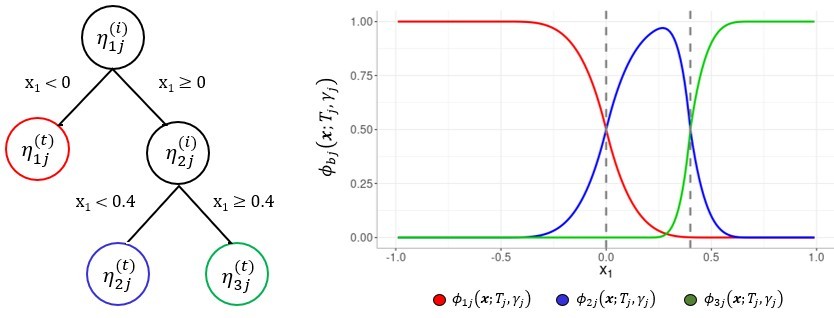}
    \caption{(Left) An example tree with $B_j = 3$ terminal nodes (red, blue, green). The internal nodes (black) define a set of splitting rules that recursively partition the input space. (Right) Example path probabilities, $\phi_{bj}(\xvec;T_j,\gamma_j)$, using $\gamma_j  = 0.5$. Each curve displays the probability of being mapped to the corresponding terminal node as a function of $x_1$. 
    }
\label{fig:rpath_phix_tree}
\end{figure}

The path probabilities, $\phi_{bj}(\xvec;T_j,\gamma_j)$, can be defined in terms of the individual splits at each internal node. Consider the tree with $B_j = 3$ terminal nodes in Figure \ref{fig:rpath_phix_tree}. The first internal node, $\eta_{1j}^{(i)}$, splits using $v_{1j} = 1$ and $c_{1j} = 0$, while the second internal node, $\eta_{2j}^{(i)}$, splits using $v_{2j} = 1$ and $c_{2j} = 0.4$. The probability of reaching $\eta_{1j}^{(t)}$ is simply the probability  of splitting left at $\eta_{1j}^{(i)}$, which is given by
    $\phi_{1j}(\xvec;T_j,\gamma_j) = 1- \psi(\xvec;1,0,\gamma_j).$
Meanwhile, an observation is mapped to the blue terminal node by first splitting right at $\eta_{1j}^{(i)}$ and then left at $\eta_{2j}^{(i)}$. This occurs with probability
    $\phi_{2j}(\xvec;T_j,\gamma_j) = \psi(\xvec;1,0,\gamma_j)\times \big(1- \psi(\xvec;1,0.4,\gamma_j)\big).$
Figure \ref{fig:rpath_phix_tree} displays the resulting path probabilities for each terminal node with $\gamma_j = 0.5$. 

In general, if the path from the root node to the $\bth$ terminal node depends on $D$ internal nodes, the path probability is defined by 
\begin{equation}
    \phi_{bj}(\xvec;T_j,\gamma_j) = \prod_{d=1}^D \psi(\xvec;v_{(dj)},c_{(dj)},\gamma_j)^{R_{(dj)}} \times \big(1-\psi(\xvec;v_{(dj)}, c_{(dj)},\gamma_j)\big)^{1-R_{(dj)}}, \label{phixdef} 
\end{equation}
where $v_{(dj)}$ and $c_{(dj)}$ are the variable and cutpoint selected at the $\dth$ internal node along the specified path in $T_j$, and $R_{(dj)} = 1$ ($R_{(dj)} = 0$) if a right (left) move  is required at the $\dth$ internal node to continue along the path towards $\eta^{(t)}_{bj}$.  

\subsection{Smooth Mean Predictions} \label{subsect:smooth_mp}

Let $Y(\tilde{\xvec})$ be a new observable quantity at input $\tilde{\xvec}$. Conditional on the $m$ random path vectors at $\tilde{x}$, the mean function is given by $$E[Y(\tilde{\xvec}) \mid \{T_j,M_j,\tilde{Z}_j,\gamma_j\}_{j = 1}^m] = \sum_{j = 1}^m \sum_{b = 1}^{B_j} \muvec_{bj}\;z_{bj}(\tilde{\xvec}),$$
where $\tilde{Z}_j = Z_j \cup \{ \zvec_j(\tilde{\xvec})\}$ is the set of random path assignments from the training data and the future observation. Marginalizing over $\tilde{Z}_j$ results in a smooth mean prediction,
\begin{align}
    E[Y(\tilde{\xvec}) \mid \{T_j,M_j,\gamma_j\}_{j = 1}^m] &=  \sum_{j = 1}^m \sum_{b = 1}^{B_j} \muvec_{bj}\;\phi_{bj}(\tilde{\xvec}; T_j, \gamma_j). \label{rpath_mean_pred}
\end{align}
Posterior samples of this expectation can be obtained by evaluating the functional form in (\ref{rpath_mean_pred}) given posterior draws of $T_j$, $M_j$, and $\gamma_j$.   

\subsection{The Semivariogram} \label{subsect:rpath_semivg}

The RPBART model introduces an additional pair of hyperparameters, $\alpha_1$ and $\alpha_2$, which control the $m$ bandwidth parameters $\gamma_j,$ and therefore the level of smoothing in the model. The common approach in BART is to calibrate the prior hyperparameters using a lightly data-informed approach or cross validation. However, neither the data-informed approach proposed for BART \citep{bart_2010} nor the default prior settings in SBART \citep{linero_sbart} can be used for calibrating the new 
smoothness parameters $\alpha_1$ and $\alpha_2$. Furthermore, the added complexity of RPBART would render the traditional cross-validation study too complex and computationally expensive. 

We propose to calibrate the RPBART model using the semivariogram \citep{cressie2015statistics}. The semivariogram, $\overline{\nu}(\lVert \hvec \rVert)$, is defined by \begin{align}
    \overline{\nu}(\lVert \hvec \rVert) = \frac{1}{|\X|} \int_\X 
    \nu(\xvec, \hvec) \; d\xvec, \label{integral_svg_var} \\[5pt]
    \nu(\xvec, \hvec) = \frac{1}{2} \text{Var}\big(Y(\xvec+\hvec)-Y(\xvec)\big), \label{nu_xh}
\end{align} 
where $|\X|$ denotes the volume of the input domain $\X$ and $\xvec+\hvec$ denotes an input which is a distance $\lVert \hvec\rVert>0$ away from $\xvec$ \citep{matheron1963principles}. Assuming a constant mean for $Y(\xvec)$, the function $\nu(\xvec, \hvec)$ simplifies as
$\nu(\xvec, \hvec) = \frac{1}{2}E\Big[\big(Y(\xvec+\hvec)-Y(\xvec)\big)^2 \Big]$. The function $\nu(\xvec, \hvec)$ describes the spatial correlation between two points that are separated by a distance of $\lVert \hvec \rVert$ and may depend on parameters that determine the shape of the semivariogram $\overline{\nu}(\lVert \hvec \rVert)$.  

We propose calibrating the hyperparameters of the RPBART model by using an estimator of the prior semivariogram in (\ref{integral_svg_var}) and the function $\nu(\xvec,\hvec)$ in (\ref{nu_xh}). The estimator of (\ref{nu_xh}) can be calculated by marginalizing  over the set of parameters $\lbrace T_j, M_j, Z_j, \gamma_j \rbrace_{j = 1}^m$ and conditioning on $\sigma^2$. In order to compute $\nu(\xvec, \hvec)$, we first analytically marginalize over $\lbrace M_j, Z_j \rbrace_{j = 1}^m$ conditional on the remaining parameters. An expression for $\nu(\xvec, \hvec)$ is then computed by numerically integrating over $\lbrace T_j, \gamma_j \rbrace_{j = 1}^m$ (where $\sigma^2$ is held fixed). The function $\nu(\xvec, \hvec)$ for the RPBART model is given by Theorem \ref{thm_psvg}.

\begin{theorem_label} \label{thm_psvg}
Assume the random quantities $\lbrace T_j, M_j, Z_j, \gamma_j \rbrace_{j = 1}^m$ are distributed as specified in Section \ref{subsect:rpath_prior}. Conditional on $\sigma^2$, the function $\nu(\xvec, \hvec)$ for the RPBART model is  
\begin{align}
    \nu(\xvec, \hvec)
    &=  \sigma^2 + m\tau^2\Big(1 - \bar{\Phi}(\xvec,\hvec)\Big), \label{thm_rpath_uncond_svg}
\end{align}
where $m\tau^2 = \Big(\frac{y_\text{max}-y_\text{min}}{2k}\Big)^2$ defines the variance of the sum-of-trees, $k$ is a tuning parameter, $y_\text{max} - y_\text{min}$ is the range of the observed data, and
\begin{align*}
    \bar{\Phi}(\xvec,\hvec)= E\Big[\sum_{b=1}^{B_1} \phi_{b1}(\xvec+\hvec;\gamma_1, T_1)\phi_{b1}(\xvec;\gamma_1,T_1)\Big]
\end{align*}
is the probability that two observations with inputs $\xvec$ and $\xvec+\hvec$ are assigned to the same partition. Without loss of generality, the expectation is with respect to $T_1$ and $\gamma_1$ (since the $m$ trees are a priori i.i.d.).
\end{theorem_label}
The proof of Theorem \ref{thm_psvg} is in Section \ref{subsect:proof_rpath_svg} of the Supplement. The expectation in Theorem \ref{thm_psvg}, $\bar{\Phi}(\xvec,\hvec)$, can be approximated using draws from the prior.  Thus, $\nu(\xvec,\hvec)$ depends on the probability that two points are assigned to different partitions, averaged over the set of trees and bandwidths, as denoted by $1 - \bar{\Phi}(\xvec,\hvec)$. This probability is then scaled and shifted by the variance of the sum-of-trees, $m\tau^2$, and error variance, $\sigma^2$. 

Generally, we are more interested in $\overline{\nu}(\lVert \hvec \rVert)$, which describes the variability across the entire input space rather than at a specific $\xvec$. For RPBART, we can numerically compute $\overline{\nu}(\lVert \hvec \rVert)$ by integrating (\ref{thm_rpath_uncond_svg}) over $\X$ as shown in (\ref{integral_svg_var}). We can use the resulting a priori semivariogram estimator to guide the selection of the hyperparameters, $\alpha$, $\beta$, $\alpha_1$, $\alpha_2$, and $k$, across the various priors in the model. Note the number of trees $m$ must still be selected through other means, such as cross-validation.

We note this procedure for computing the semivariogram is more complex than what is observed with Gaussian Processes \citep{cressie2015statistics}. In such cases, shift-invariant kernels are typically selected for the covariance model in a stationary Gaussian Process. In other words, the covariance model is simply a function of $\hvec$ under a shift-invariant kernel. However, the RPBART covariance function, $\bar{\Phi}(\xvec,\hvec)$, is a non-shift-invariant product kernel of the path probabilities that is averaged over all possible trees and bandwidth parameters. Because of this, we must consider the computation of $\nu(\xvec, \hvec)$ in order to numerically compute $\overline{\nu}(\lVert \hvec \rVert)$.

\begin{figure}[t]
    \centering
    \includegraphics[width = 1\textwidth, height = 0.375\textwidth]{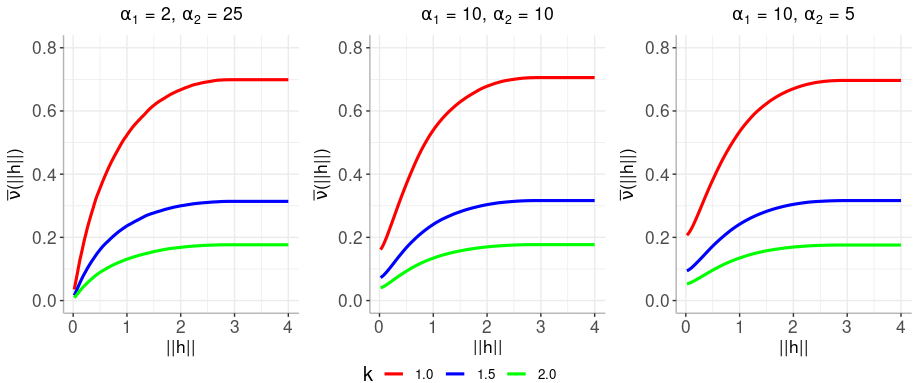}
    \caption{The semivariogram for different values of $k = 1$ (red), $k = 1.5$ (blue), and $k= 2$ (green) and bandwidth hyperparameters $\alpha_1$ and $\alpha_2$. Each semivariogram is generated using $y_{min} = -1$, $y_{max} = 1$, $\sigma = 0$, and $\alpha = 0.95$, and $\beta = 0.5$.}
\label{fig:svg_deep_trees}
\end{figure}

For example, consider the semivariogram with respect to the 2-dimensional input space $[-1,1]\times [-1,1]$. 
Figure \ref{fig:svg_deep_trees} displays possible semivariograms for different values of $k$, $\alpha_1$, and $\alpha_2$ with $y_{min} = -1$ and $y_{max} = 1$. Note, we set the nugget $\sigma = 0$ in this example to clearly isolate the effect of the smoothness parameters $\alpha_1$ and $\alpha_2$, i.e. ignoring the nugget. 
Each panel corresponds to different settings of the bandwidth prior, where $\alpha_1=2$ and $\alpha_2=25$ indicate low levels of smoothing and 
$\alpha_1=10$ and $\alpha_2=5$ indicate high levels of smoothing. Based on each curve, we see the smoothing parameters primarily affect the behavior of the semivariogram for small values of $\lVert \hvec \rVert$. With less smoothing (left), each of the semivariograms take values closer to $0$
when $\lVert \hvec \rVert$ is small, while a noticeable shift upwards is observed with more smoothing (center, right). This offset appears in the RPBART semivariogram regardless of the $\gamma_j$ prior because the covariance function is discontinuous. 

As $\lVert \hvec \rVert$ increases, each semivariogram reaches a maximum value, known as the sill. For each $k$, we observe the sill is similar regardless of the values of $\alpha_1$
and $\alpha_2$. Thus, the value of $k$ controls the height of the sill and in turn the amount of variability attributed to the sum-of-trees model. As a result, the interpretation of $k$ under the RPBART model is the same as in the original BART model. Finally, we should note that the hyperparameters in the tree prior, $\alpha$ and $\beta$, can also be determined based on the semivariogram, as deeper trees result in a larger number of partitions and 
thus less correlation across the input space. Hence, deeper trees shift the semivariograms upwards and further contribute to the offset. 

In practice, we can compare the theoretical semivariogram to the empirical semivariogram to select the hyperparameters for the model \citep{cressie2015statistics}. In summary, we see $\alpha_1$ and $\alpha_2$ primarily affect the semivariogram when $\lVert \hvec \rVert$ is near $0$, as more smoothing generally leads to an upward shift and increase in curvature near the origin. The tree prior hyperparameters, $\alpha$ and $\beta$, affect the size of the offset and the height of the sill. Meanwhile, the value of $k$ simply scales the function and has minimal impact on the overall shape of the curve. 

The last component to consider is the value of $\sigma^2$, which is set to $0$ in Figure \ref{fig:svg_deep_trees}. The typical way to calibrate the $\nu\lambda/\chi_\nu^2$ prior on $\sigma^2$ is to fix $\nu$ at a desired value and select $\lambda$ based on an estimate of $\sigma^2$, denoted by $\hat{\sigma}^2$ \citep{bart_2010,linero_sbart}. Traditional approaches typically compute $\hat{\sigma}^2$ using a linear model. From Theorem \ref{thm_psvg}, we see $\sigma^2$ simply adds to the offset by vertically shifting the semivariogram. Thus, we can select $\hat{\sigma}^2$ based on theoretical and empirical semivariogram. Given a value of $\nu$, one can set $\hat{\sigma}^2$ as the mode of the prior distribution and algebraically solve for $\lambda$. This strategy selects $\lambda$ along with the other hyperparameters using the information encoded in the semivariogram. 

Additional examples of the RPBART semivariogram, including a real world application to climate models, can be found in the supplementary material.

\section{Smooth Model Mixing} \label{subsect:rpath_mm}

\subsection{A Mean-Mixing Approach}

The RPBART model easily extends to the model mixing framework presented by \cite{yannotty2023model}. Let $f_1(\xvec_i),\ldots,f_K(\xvec_i)$ denote the output from $K$ simulators at $\xvec_i$. When the simulators are computationally expensive, the output $f_l(\xvec_i)$ is replaced with the prediction from an inexpensive emulator, $\hat{f}_l(\xvec_i)$, for $l = 1,\ldots,K$. For example, $\hat{f}_l(\xvec_i)$ could be the mean prediction from a Gaussian process \citep{santner2018design} or an RPBART emulator. In climate applications, each GCM may be evaluated on different latitude and longitude grids, thus we use regridding techniques such as bilinear interpolation to compute $\hat{f}_l(\xvec_i)$.   

Given the mean predictions from $K$ emulators, we assume $Y(\xvec_i)$ is modeled by
\begin{align*}
    Y(\xvec_i) \mid \fhatvec(\xvec_i), \lbrace T_j, M_j, Z_j \rbrace_{j = 1}^m, \sigma^2 \; &\sim \; N\big(\fhatvec^\top(
    \xvec_i) \wvec(\xvec_i),\sigma^2), \\[4pt]
    \wvec(\xvec_i) &= \sum_{j =1}^m \sum_{b = 1}^{B_j} \muvec_{bj}z_{bj}(\xvec_i),
\end{align*}
where $\fhatvec(\xvec_i)$ is the $K$-dimensional vector of mean predictions at input $\xvec_i$, $\wvec(\xvec_i)$ is the corresponding $K$-dimensional weight vector, and $i=1,\ldots,n$. This extension allows the $K$ weights to be modeled as continuous functions using similar arguments as outlined for the 1-dimensional tree output case in Section \ref{subsect:smooth_mp}. 

Similar to \cite{yannotty2023model}, we  regularize the weight functions via a prior on the terminal node parameters. The primary goal is to ensure each weight function, $w_l(\xvec)$, prefers the interval $[0,1]$ without directly imposing a strict non-negativity or sum-to-one constraint. Thus, we assume $    \muvec_{bj} \mid T_j \sim N\Big(\frac{1}{mK} \onevec_K\;, \tau^2 I_K \Big),$
where $\tau = 1/(2k\sqrt{m})$ and $k$ is a tuning parameter. This choice of $\tau$ follows \cite{bart_2010} in that the confidence interval for the sum-of-trees has a length of $w_{max} - w_{min}$, where 0 and 1 are the target minimum and maximum values of the weights. This prior calibration ensures each $w_l(\xvec)$ is centered about $1/K$, implying each model is equally weighted at each $\xvec$ apriori. The value of $k$ controls the flexibility of the weight functions. Small $k$ allows for flexible weights that can easily vary beyond the target bounds of $0$ and $1$ and thus identify granular patterns in the data. Larger $k$ will keep the weights near the simple average of $1/K$ and could be limited in identifying regional patterns in the data. Typically, more flexible weights are needed when mixing lower fidelity models, as the weights will account for any  discrepancy between the model set and the observed data. By default, we set $k = 1$. 

RPBART allows for independent sampling of each of the $B_j$ terminal node parameters, $\muvec_{bj}$, within $T_j$ for $j = 1,\ldots,m$. Thus, RPBART avoids the joint update required in SBART. This is rather significant, as the SBART update for the $B_j$ terminal node parameters vectors would require an inversion of a $KB_j \times KB_j$ matrix. In larger scale problems with larger $K$ or deeper trees, the repeated inversion cost of this matrix would be burdensome.   

Prior calibration of the weight functions can still be informed using the semivariogram approach. Section \ref{subsect:mm_svg} provides the formula for the RPBART-BMM semivariogram along with the derivation of the formula.

In general, RPBART-BMM model introduces an additional layer of randomness into the BART-BMM framework. In BART-BMM, final mixed solution tended to overfit the data, particularly in regions where no model under consideration provided a high-fidelity approximation of the underlying system. In such regions of the domain, BART-BMM would simply find a useful combination of the models to best fit the data. The corresponding tree models would then exhibit poor mixing within this partition of the domain. The random path concept helps alleviate some of the overfitting, as training points can be mapped to adjacent terminal nodes at each iteration of the MCMC, even if the sampled tree structure remains unchanged.

\subsection{Posterior Weight Projections} \label{subsect:wt_project}

The RPBART weights are modeled as unconstrained functions of the inputs. Alternative approaches impose constraints on the weight functions, such as a non-negativity or sum-to-one constraint \citep{hs, COSCRATO2020141, breiman_stacking}. Although constrained approaches may introduce bias into the mean of the mixed prediction, they can improve the interpretability of the weight functions. However, such constraints could increase the computational complexity of the model. For example, a simplex constraint on the RPBART-based weights would drastically change the model fitting procedure, as the conditional conjugacy through the multivariate normal prior would likely be lost. 

Rather than changing the model and estimation procedure, one alternative is to explore the desired constraints through post-processing methods which impose constraints a posteriori rather than a priori \citep{lin2014bayesian, sen2018constrained}. In this regard, we define a constrained model in terms of the original unconstrained model using Definition  \ref{defn_projection}.

\begin{definition_label} \label{defn_projection}
    Let $\wvec(\xvec) = \big(w_1(\xvec),\ldots,w_K(\xvec)\big)^\top$ be a vector of unconstrained weights and $Y(\xvec)$ be an observable quantity from the underlying physical process. Define the unconstrained and constrained models for $Y(\xvec)$ by
    \begin{align*}
        \text{(Unconstrained)}\quad Y(\xvec) &= \sum_{l=1}^K w_l(\xvec) \hat{f}_l(\xvec) + \epsilon(\xvec), \\
        \text{(Constrained)}\quad Y(\xvec) &= \sum_{l=1}^K u_l(\xvec) \hat{f}_l(\xvec) + \delta(\xvec) + \epsilon(\xvec),
    \end{align*}
where $\uvec(\xvec) = \big(u_1(\xvec),\ldots,u_K(\xvec)\big)^\top$ is the projection of $\wvec(\xvec)$ onto the constrained space, and 
\begin{equation*}
    \delta(\xvec) = \sum_{l = 1}^K \big(w_l(\xvec) - u_l(\xvec) \big)\hat{f}_l(\xvec)
\end{equation*}
denotes the estimated discrepancy between the constrained mixture of simulators and the underlying process, and $\epsilon(\xvec)$ is a random error.
\end{definition_label}
This framework allows for the interpretation of the weight functions on the desired constrained space without introducing significant computational burdens. The unconstrained and constrained models are connected through an additive discrepancy, $\delta(\xvec)$, which accounts for any potential bias introduced by the set of constraints.

Posterior samples of $\uvec(\xvec)$ can easily be obtained by projecting the posterior samples of $\wvec(\xvec)$ onto the constrained space. The specific form of the projection will depend on the desired constraints. In this work, we explore projections that enforce a simplex constraint that result in continuous constrained weight functions, are computationally cheap, and promote some level of sparsity. The simplex constraint ensures each $u_l(\xvec) \geq 0$ and $\sum_{l = 1}^K u_l(\xvec) = 1$, while also enabling a more clear interpretation of the weights, as the mixed-prediction is simply an interpolation between different model predictions. The simplex constraint can be enforced by defining $\uvec(\xvec)$ as a function of $\wvec(\xvec)$ via a softmax or penalized $L_2$ projection \citep{laha2018controllable, kong2020rankmax}. These projections each have closed form expressions and are inexpensive to compute. One can choose between these two methods (among others) based on the desirable properties in the constrained weights and discrepancy. 

The softmax is used in stacking methods such as Bayesian Hierarchical Stacking \citep{hs}.
Although the softmax function is widely used, a common criticism is the lack of sparsity in that the constrained weights will take values between 0 and 1 but never
exactly reach either bound. Thus, the softmax can shrink the effect of a given model, but each of the $K$ models will have at least some non-zero contribution to the final prediction. 
The penalized $L_2$ projection (i.e. the sparsegen-linear from \cite{laha2018controllable}) defines the weights using a thresholding function. This enables more sparsity in the constrained weights, as any $u_l(\xvec)$ can take values of exactly $0$. Both methods rely on a ``temperature parameter", $\mathcal{T}$, which can be used to control the shape of the projected surface.

\begin{figure}[t]
    \centering
    \includegraphics[width = 1\textwidth, height = 0.475\textwidth]{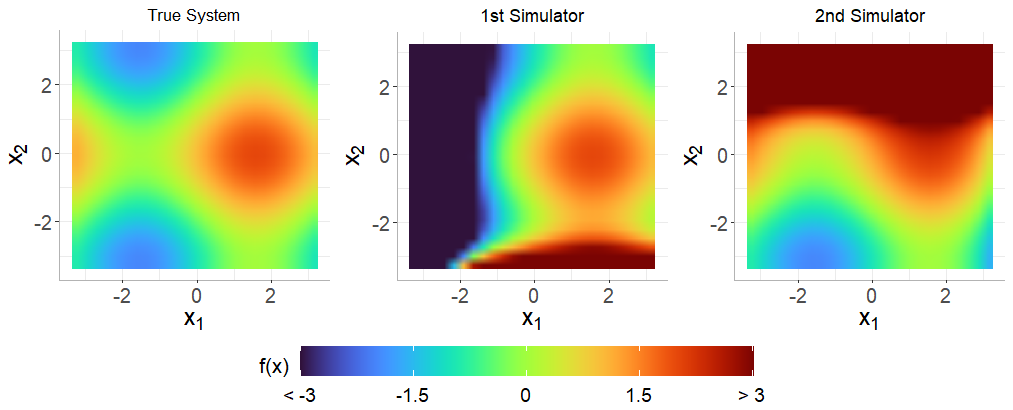}
    \caption{(Left) The underlying system $f_\dagger(\xvec)$. (Center) The output of the first simulator. (Right) The output of the second simulator.}
    \label{fig:2d_taylor_ms}
\end{figure}

\section{Applications}

This section outlines two different examples of model mixing using the RPBART method for Bayesian Model Mixing (RPBART-BMM). Section \ref{subsect:toy_ex} demonstrates the methodology on a toy simulation example, which combines $K=2$ simulators, each with two inputs. Both simulators in Section \ref{subsect:toy_ex} provide a high-fidelity approximation of the underlying system in one subregion of the domain and is less accurate across the other regions. Similar patterns are often observed in GCMs. Section \ref{subsect:climate_ex} demonstrates our methodology on a real-data application with eight GCMs that model the average monthly surface temperature.
Additional examples of RPBART and RPBART-BMM can be found in Section \ref{subsect:examples} of the supplementary material.

\subsection{A Toy Numerical Experiment} \label{subsect:toy_ex}



Consider a 2-dimensional example that mixes two simulators in the domain of $[-\pi, \pi] \times [-\pi, \pi]$. Figure \ref{fig:2d_taylor_ms} illustrates the true underlying function (right) and the two mean predictions from the simulators (center and left). The first simulator (center) provides a high-fidelity approximation of the system in the domain of $[0,\pi] \times [0 ,\pi]$. Meanwhile, the second simulator (left) provides a high-fidelity approximation when $x_2 < 0$. Neither simulator accurately predicts the true system in the top left corner of the domain. In particular, both simulators take values that are noticeably different than the underlying system.

\begin{figure}[t]
    \centering
    \includegraphics[width = 1\textwidth, height = 0.82\textwidth]{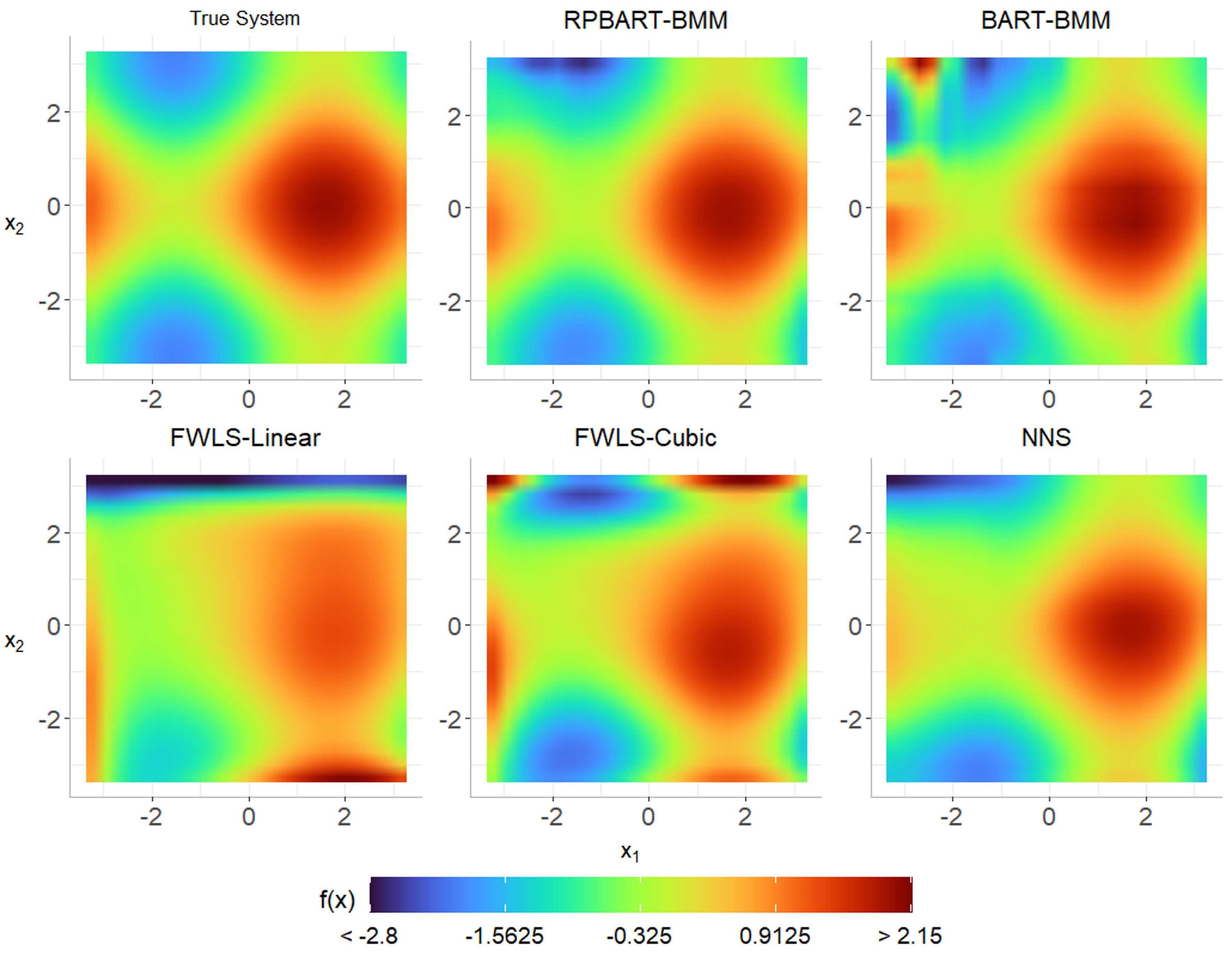}
    \caption{(Top row) The true system, $f_\dagger(\xvec)$, (left) vs. the mean predictions, $\hat{f}_\dagger(\xvec)$,  using the RPBART-BMM (center) or BART-BMM (right). (Bottom row) The mean predictions from Feature Weighted Linear Stacking (FWLS) with a linear basis (left), FWLS with a cubic basis (center), and Neural Network Stacking (NNS) (right).}
    \label{fig:2d_rpath_dpath_pred}
\end{figure}

A sample of 80 observations were independently generated from the underlying system with mean $f_\dagger(\xvec)$ and standard deviation 0.1. The BART-BMM and RPBART-BMM models were fit using this data and the final predictions are shown in Figure \ref{fig:2d_rpath_dpath_pred}.

The root mean squared prediction errors are 0.14 and 0.357 for RPBART-BMM and BART-BMM, respectively. The clear distinction between the two mean predictions lies within the top left corner of the domain, as RPBART-BMM does not overfit to the training data.  

\begin{figure}[t]
    \centering
    \includegraphics[width = 1\textwidth, height = 0.42\textwidth]{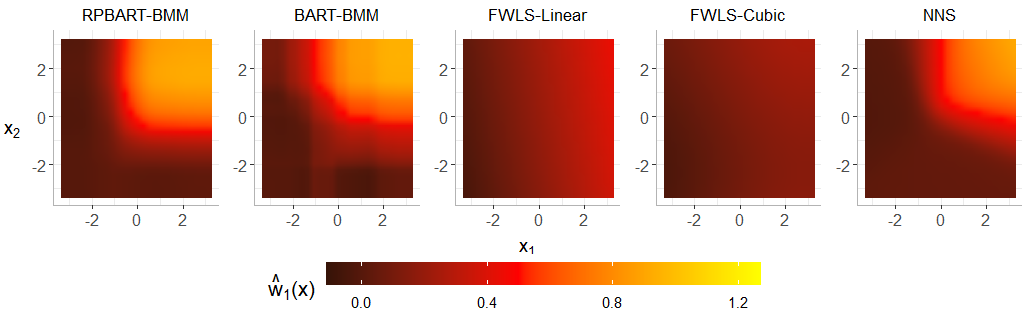}
    \caption{The mean predictions of the weight functions, $\hat{w}_1(\xvec)$ using RPBART-BMM, BART-BMM, FWLS-Linear, FWLS-Cubic, and NNS.}
    \label{fig:2d_rpath_wts_w1}
\end{figure}

Additionally, we apply Feature Weighted Linear Stacking (FWLS) \citep{sill2009feature} and  Neural Network Stacking (NNS) \citep{COSCRATO2020141}. The FWLS weight functions are modeled using pre-specified basis functions in terms of $x_1$ and $x_2$. Figure \ref{fig:2d_rpath_dpath_pred} displays the mean predictions for FWLS using a linear basis (FWLS-Linear, bottom row, left) and FWLS using a cubic basis (FWLS-Cubic, bottom row, center). Both FWLS predictions struggle to capture the distinct features of the system, particularly around the edges of the domain. In both cases, we observe how pre-specified basis functions may be insufficient for recovering the underlying system across the entire domain. The neural network stacking model (3 hidden layers with 100 nodes per layer) performs better than the linear model, although it also struggles to capture the function behavior around the domain.     

\begin{figure}[t]
    \centering
    \includegraphics[width = 1\textwidth, height = 0.42\textwidth]{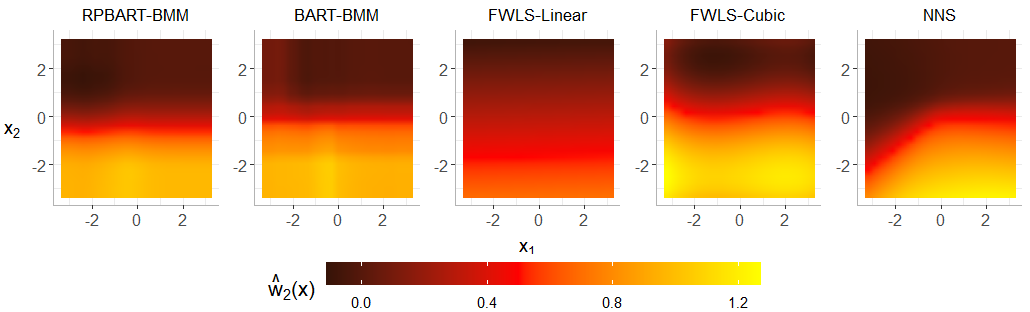}
    \caption{(Top) The mean predictions of the weight functions, $\hat{w}_2(\xvec)$ using RPBART-BMM, BART-BMM, FWLS-Linear, FWLS-Cubic, and NNS.}
    \label{fig:2d_rpath_wts_w2}
\end{figure}

The mean weight functions for the first simulator are shown in Figure \ref{fig:2d_rpath_wts_w1}, while the mean weight functions for the second simulator are displayed in Figure \ref{fig:2d_rpath_wts_w2}. Overall, the underlying inference is very similar across the two BART-based methods and NNS. The first simulator receives weight near 1 in the top right corner of the domain where it is the best approximator, the second simulator receives weight near 1 in the bottom half of the domain, and a useful combination of the simulators is determined in the top left corner. Meanwhile, inference using the two FWLS methods is less clear. FWLS-Linear is not flexible enough to produce meaningful inference, as the weights exhibit subtle changes over the domain and neither model is heavily favored over the other. The added complexity of the cubic basis allows FWLS-Cubic to yield more reliable inference, particularly with the second weight function.

\subsection{Application to Climate Data Integration} \label{subsect:climate_ex}

In this application, we mix multiple GCMs which model the monthly average two-meter surface temperature (T2M) across the world for April, August, and December 2014. These three time periods are chosen to capture how the GCM performance varies across different months. We combine the output of eight different simulators, each with varying fidelity across the input space and spatial resolution. 

We compare the RPBART-BMM approach to Feature-Weighted Linear Stacking and Neural Network Stacking. We allow all three methods to model the weights as functions of latitude, longitude, elevation, and month. Each mixing method is trained using 45,000 observations, where 15,000 are taken from each of the three time periods. The predictions for April, August, and December 2014 are generated over a grid of 259,200 latitude and longitude pairs for each time period.   


We downloaded the GCM data from the Coupled Model Intercomparison Project \\ (CMIP6) \citep{eyring2016overview}, a data product which includes outputs from a wide range of simulators used to study various climate features. Each GCM may be constructed on a different set of external forcing or socioeconomic factors, hence the fidelity of each climate model is likely different across the globe. Regardless of these factors, each GCM outputs the T2M on a longitude and latitude grid although the grid resolutions can be different. We denote the output from each GCM at a given input $\xvec$  as $f_l(\xvec)$ where $l=1,\ldots,K$.

\begin{table}[t]
\centering
\begin{singlespace} 
\begin{tabular}{ |p{3cm}||c|c|c|  }
 \hline
 Model&April 2014& August 2014 &December 2014\\
 \hline
 RPBART-BMM   & {\bf 0.827}    & {\bf 0.864} &   {\bf 0.882}\\
 NNS   & 0.993    &1.032&  1.088 \\
 FWLS   & 2.393    &2.345 & 2.142\\
 \hline
\end{tabular} 
 \caption{The root mean square error for each mixing approach (BART-BMM, NNS, and FWLS) stratified by month. Each RMSE is computed by evaluating the output of the mixed-prediction or a given GCM over a dense grid of 259,200 inputs and comparing to the observed ERA5 data.  Smallest RMSEs for each month are denoted in bold.}
 \label{tbl:world8_rmse}
\end{singlespace}
\end{table}

\begin{figure}[t]
    \centering
    \includegraphics[width = 1\textwidth, height = 0.89\textwidth]{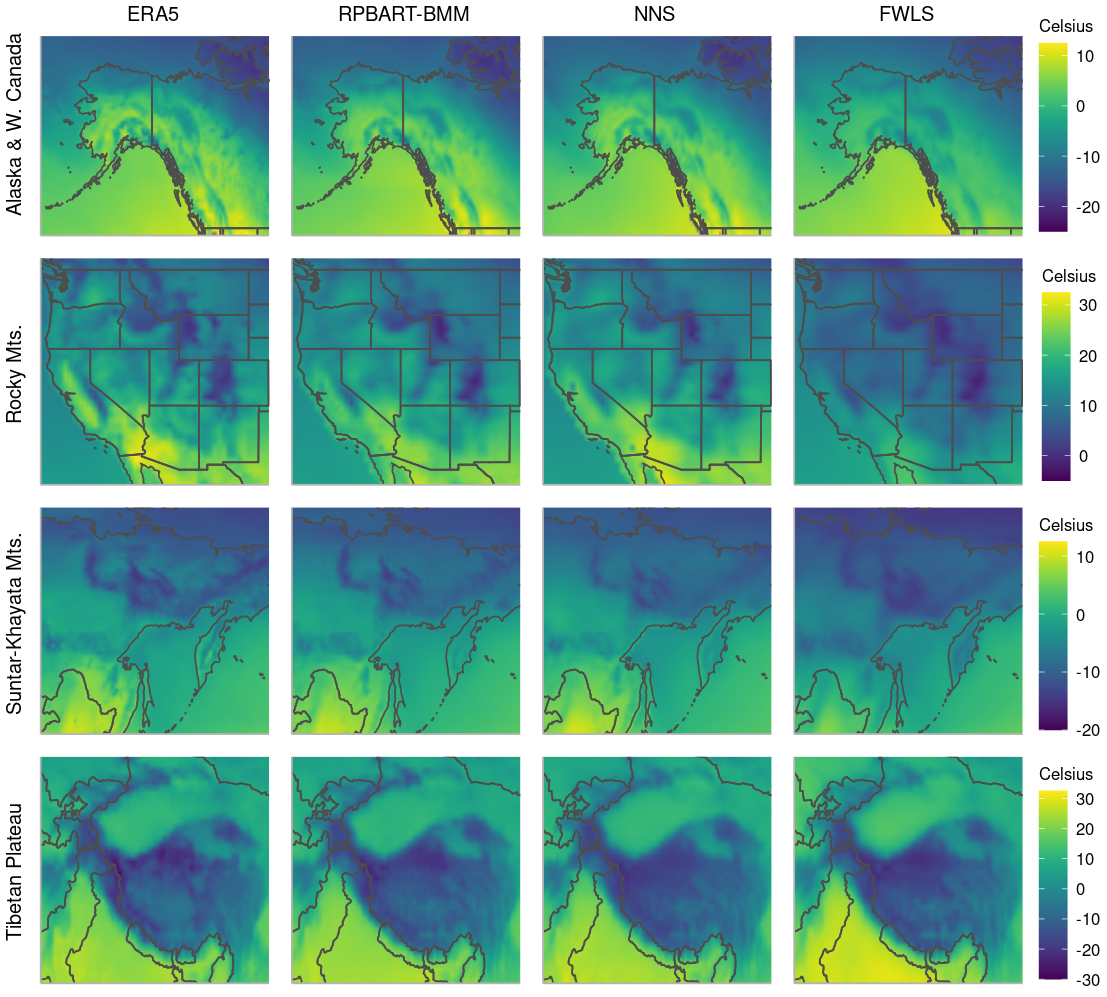}
    \caption{The ERA5 data (first column) vs the mean predictions from RPBART-BMM (second column), NNS (third column), or FWLS (fourth column). Each row highlights different regions of the world in April 2014. }
\label{fig:world_mixing8_pred}
\end{figure}

In climate applications, reanalysis data is often used for observational data. We obtain the European Centre for Medium-Range Forecasts reanalysis (ERA5) data, which combines observed surface temperatures with results from a weather forecasting model to produce measurements of T2M across a dense grid of $0.25^{\circ}$ longitude by  $0.25^{\circ}$ latitude. We denote the $\ith$ reanalysis data point as $Y(\xvec_i)$ where $\xvec_i$ denotes the $\ith$ latitude, longitude, elevation, and month for $i=1,\ldots,45000$. The ERA5 data and elevation data was downloaded from the Copernicus Climate Data Store.

A common assumption in model mixing is that the simulators are evaluated across the same grid of inputs as the observed response data. 
To map the data onto the same grid, we apply the  bilinear interpolation to obtain an inexpensive emulator, $\hat{f}_1(\xvec),\ldots,\hat{f}_K(\xvec),$ for each GCM \citep{ncar_regrid}. 


Table \ref{tbl:world8_rmse} displays the root mean squared errors for each of the three time periods. The RPBART-BMM model performs the best across each month. We also note each of the three mixing methods outperform the individual GCMs as shown in Section \ref{gcm_table} of the Supplementary Material. Figure \ref{fig:world_mixing8_pred} displays the ERA5 data (left) and the mean predictions from the three approaches across 
four different regions in April 2014. In general, RPBART-BMM and NNS result in very similar predictions that capture the granular features in the data. Some of the granularity is lost in FWLS due to the less flexible form of the weight functions defined by a linear basis. Only subtle differences exist between RPBART-BMM and NNS. Also, RPBART-BMM better leverages elevation to preserve the fine details in the data, as seen across the Suntar-Khayata Mountains and the Tibetan Plateau predictions.   

\begin{figure}[h]
    \centering
    \includegraphics[width = 1\textwidth, height = 0.62\textwidth]{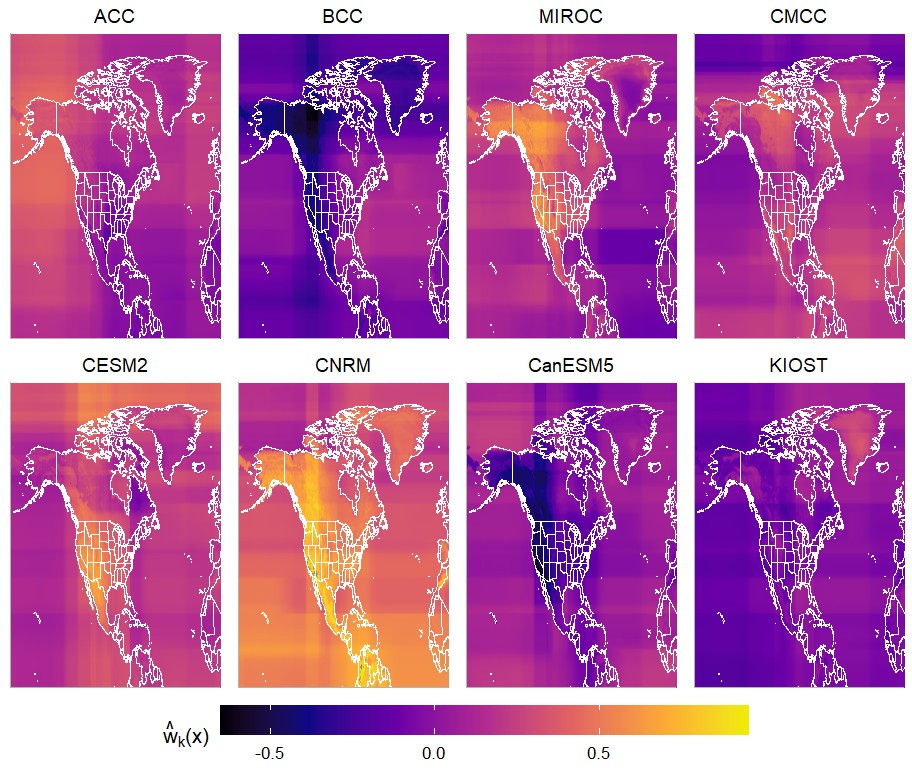}
    \caption{The posterior mean of the RPBART-based weight functions for each of the eight simulators across the northwestern hemisphere in April 2014.}
\label{fig:world_mixing8_wts_nwh}
\end{figure}

In addition to the improved mean prediction, we are also interested in identifying the subregions where each simulator is favored in the ensemble. In most cases, we have observed that weight values closer to $1$ typically indicates more accurate GCMs, while weights near $0$ typically indicate less accurate GCMs. Figure \ref{fig:world_mixing8_wts_nwh} displays the posterior mean weight functions for the eight simulators in the Northwestern hemisphere in April 2014. The posterior mean weights in Figure \ref{fig:world_mixing8_wts_nwh} suggest CNRM, CESM2, and MIROC are more active in the mixed-prediction across the western part of the United States and Canada, as these three GCMs receive higher weight relative to the other GCMs. In the same region, we see BCC and CanESM5 receive negative weights, which are possibly due to multicollinearity and can be investigated using posterior weight projections as described in Section \ref{subsect:wt_project}. 

\begin{figure}[t]
    \centering
    \includegraphics[width = 1\textwidth, height = 0.67\textwidth]{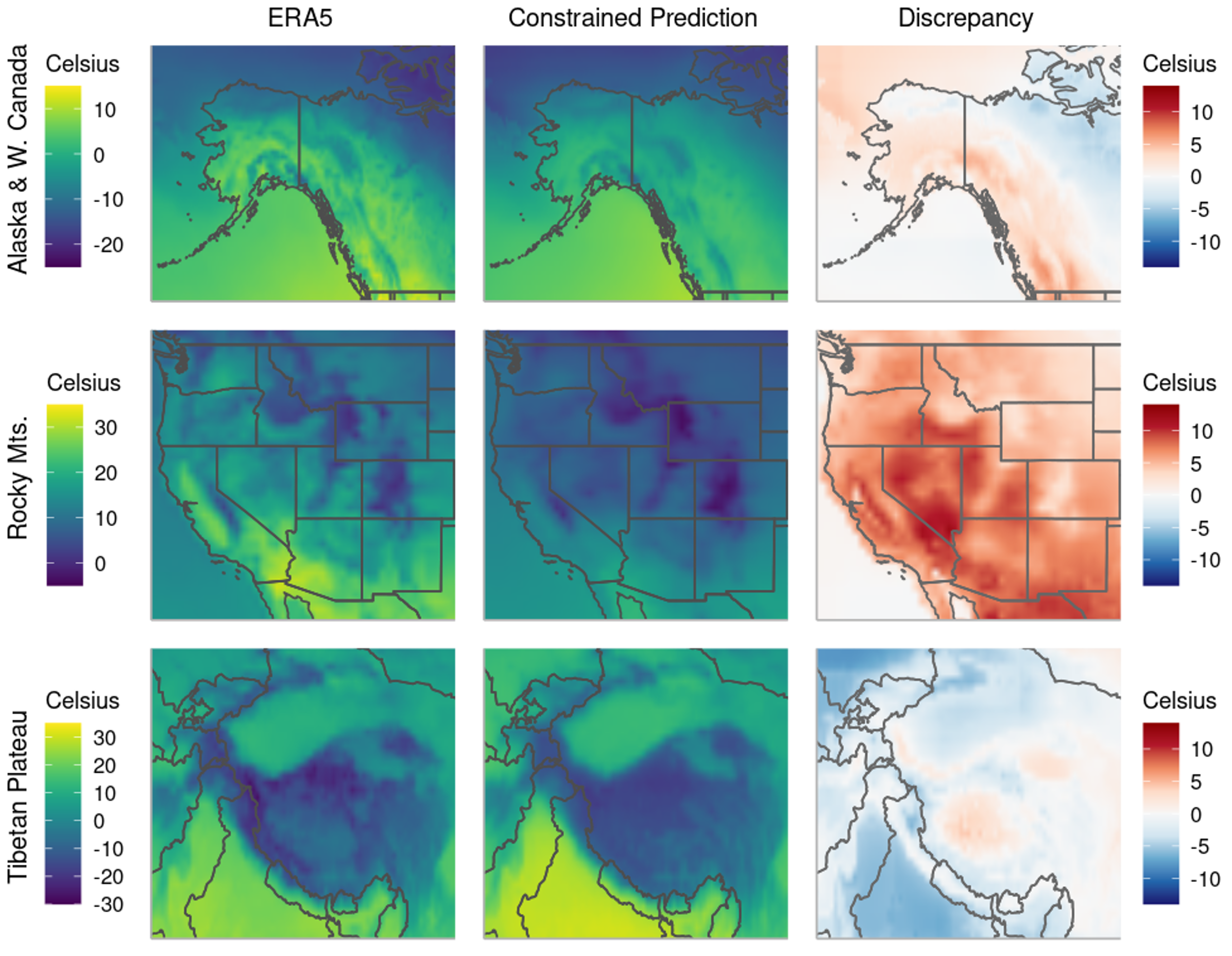}
    \caption{The ERA5 data (left) vs. the simplex constrained mixture of simulators (center) and the additive discrepancy (right) across three subregions (rows) in April 2014. }
\label{fig:world_mixing8_proj_delta}
\end{figure}

\begin{figure}[h]
    \centering
    \includegraphics[width = 1\textwidth, height = 0.58\textwidth]{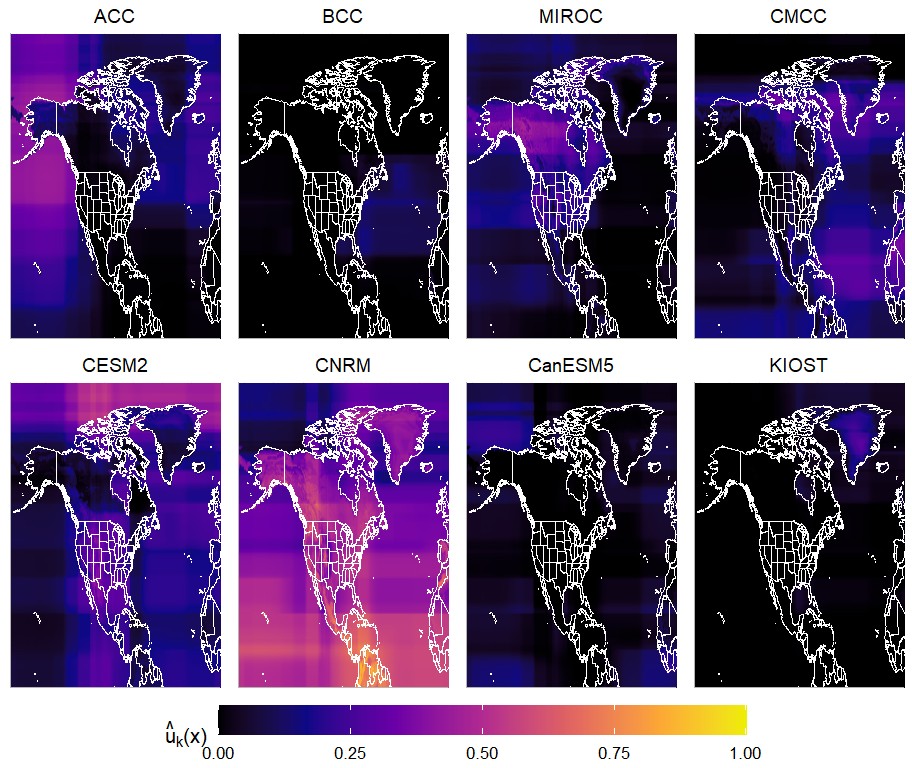}
    \caption{The posterior mean of the constrained RPBART-based weight functions for each of the eight simulators across the northwestern hemisphere in April 2014.}
\label{fig:world_mixing8_cwts_nwh}
\end{figure}

To better identify a subset of GCMs that are active in a specific region, we consider projecting the samples of the posterior weight functions onto the simplex using a penalized $L_2$ projection (the sparsegen-linear projection) \citep{laha2018controllable}. 
This imposes a sum-to-one and non-negativity constraint on $\uvec(\xvec)$. For this example $\mathcal{T} = 0.137$ by minimizing the sum of squared differences between the predicted mean with unconstrained weights and the predicted mean with the constrained weights (this difference is defined as the discrepancy function), $\sum_{i = 1}^{N_v} \big(\delta(\xvec^v_i)\big)^2$, at $N_v = 5000$ validation points.

The posterior mean predictions from the constrained mixture of GCMs and the posterior mean discrepancy are shown in Figure \ref{fig:world_mixing8_proj_delta}. Overall, the constrained mixed-prediction recovers the major features of the underlying temperature patterns, however some fine details are lost due to the simplex constraint. This, at times, can be indicative of response features that are not accounted for by any of the GCMs in the model set. Typically, we lose some of the granularity in the areas with significant elevation changes, such as the Rocky Mountains (second row). The constrained weights are less flexible and unable to properly account for the discrepancy within the GCMs. The remaining variability that is unaccounted for by the constrained mixed-prediction is then attributed to the additive discrepancy.

Figure \ref{fig:world_mixing8_cwts_nwh} displays the projected weight functions for April 2014. Similar conclusions can be made as in the unconstrained weights from Figure \ref{fig:world_mixing8_wts_nwh}, however in the constrained model, we can clearly isolate the GCMs that contribute to the prediction in Figure \ref{fig:world_mixing8_proj_delta}. For example, we see CNRM, CESM2, and MIROC receive relatively high weight in the western part of the United States and Canada. Coupled with the discrepancy in Figure  \ref{fig:world_mixing8_proj_delta}, these three GCMs are sufficient for recovering the system in Alaska and Western Canada (low $\delta(\xvec)$), but are insufficient for recovering the system in the Rocky Mountains (higher $\delta(\xvec)$). 

\begin{figure}[h]
    \centering
    \includegraphics[width = 1\textwidth, height = 0.64\textwidth]{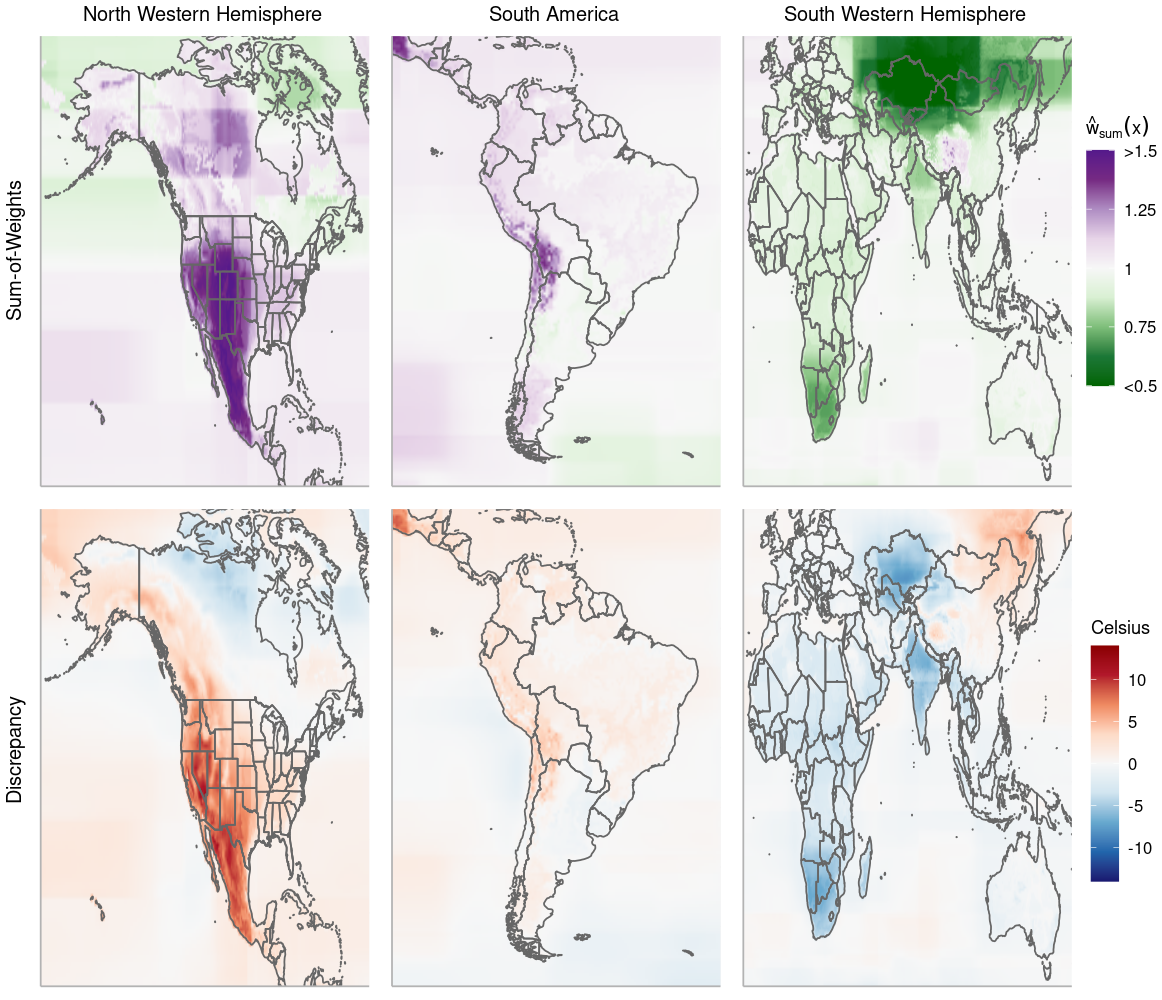}
    \caption{The sum-of-weights from the unconstrained RPBART-BMM model (top) and the additive discrepancy from the constrained weight model (bottom). Each column corresponds to a different region in April 2014.}
\label{fig:world_mixing8_sum_wts_delta}
\end{figure}

Finally, we can connect the unconstrained and constrained models in terms of how each identifies model discrepancy. In the unconstrained model, the posterior sum of the weight functions, $w_{sum}(\xvec) = \sum_{l = 1}^K w_l(\xvec)$, is one metric that can be used to help identify model discrepancy. In our applications (both climate and nuclear physics), we have observed that the posterior distribution of $w_{sum}(\xvec)$ deviates away from $1$ in areas where the GCMs do not accurately model the underlying system. This phenomenon has also been observed in earlier work  with constant weights \citep{breiman_stacking,le2017bayes}. 
The top panel of Figure \ref{fig:world_mixing8_sum_wts_delta} displays the posterior mean of $w_{sum}(\xvec)$ over selected regions in April 2014. The bottom row of Figure \ref{fig:world_mixing8_sum_wts_delta} displays the posterior mean of $\delta(\xvec)$ from the constrained model in April 2014. Subregions where $w_{sum}(\xvec)$ deviates away from $1$ (purple or green) align well with the areas where $\delta(\xvec)$ deviates away from $0$ (orange or blue). Thus, the post-processing approach used to estimate $\delta(\xvec)$ preserves the information in the posterior of $w_{sum}(\xvec)$. 

\section{Discussion}

This  research  proposes a random path model as a novel approach to introduce continuity into the sum-of-trees model. We then extend the original mean mixing framework presented by \cite{yannotty2023model} to combine the outputs from a general collection of GCMs while also modeling the weight functions as continuous functions. This methodology has been successfully applied to GCMs, which model the monthly average surface temperature. 

This model mixing approach depends on information from the observational data and all GCM outputs. Our method explicitly assumes one can evaluate each GCM, or an emulator for the GCM, at the $n$ inputs associated with the  observational data. We construct each emulator using bilinear interpolation with respect to the latitude and longitude grid associated with the monthly temperature output. This results in a simple, yet lower resolution emulator for each GCM. Since the emulators maintain a lower resolution, the weight functions in the RPBART-BMM model are left to account for the granularity in the observed data. Thus, lower resolution emulators require more complex weight functions. For RPBART-BMM, the required complexity corresponds to deep trees, less smoothing, and larger $m$. 

Alternatively, one could replace the bilinear interpolation approach with a more complex emulator, such as another RPBART model, neural networks, or Gaussian processes \citep{watson2021machine, kasim2021building}. Such emulators could depend on more features other than longitude and latitude and thus result in higher fidelity interpolations of the GCM output. Higher resolution emulators would likely explain more granularity in the data and thus result in less complex  weight functions (i.e. shallower trees, more smoothing, and smaller $m$).  

Existing climate ensembles tend to focus on the temporal distribution of the simulator output and combine the GCMs in a pointwise manner for each latitude and longitude pair \citep{vrac2024distribution, harris2023multimodel}. Though a temporal component can be considered as an input to the weight functions, we mainly focus on learning the spatial distribution of the weights to help identify subregions where each GCM is influential. Future work will further explore other ways to incorporate the temporal distribution of the GCMs, which could allow for more appropriate long term model-mixed predictions of future temperature.

\section{Acknowledgements}
The work of JCY was supported in part by the National Science Foundation under Agreement OAC-2004601.
The work of MTP was supported in part by the National Science Foundation under Agreements DMS-1916231, DMS-1564395, OAC-2004601.
The work of TJS was supported in part by the National Science Foundation under Agreement DMS-1564395 (The Ohio State University).
The work of BL was supported in part by the National Science Foundation under Agreement DMS-2503494.

\begin{singlespace} 
\bibliography{references}{}
\end{singlespace}

\clearpage

    
\section*{Supplement}

\subsection{The Conditional Covariance Model}

We can better understand the parameters in the RPBART model by studying their effects on the covariance of the response for two inputs, $\xvec$ and $\xvec^\prime$. Conditional on the set of trees, $T_1,\ldots,T_m$, bandwidth parameters $\gamma_1,\ldots,\gamma_m$, and error variance $\sigma^2$, the prior covariance between $Y(\xvec)$ and $Y(\xvec^\prime)$ is given by Theorem \ref{thm:ccv}.

\begin{theorem_label}[Conditional Covariance] \label{thm:ccv}
Assume the $m$ sets of terminal node parameters, \\ $M_1,\ldots,M_m$, and random path assignments, $Z_1,\ldots,Z_m$ are mutually independent conditional on the set of trees and bandwidth parameters. Further assume that the random path assignments for an observation $\xvec$ are conditionally independent of those for another input $\xvec^\prime$. Then, conditional on the set of trees, $T_1,\ldots,T_m$, bandwidth parameters $\gamma_1,\ldots,\gamma_m$, and error variance $\sigma^2$ the prior covariance between $Y(\xvec)$ and $Y(\xvec^\prime)$ when $\xvec \ne \xvec^\prime$ is given by   
\begin{equation} 
\text{Cov}\big(Y(\xvec),\;Y(\xvec^\prime)\mid \Theta \big) = m\tau^2 \sum_{j = 1}^m \frac{1}{m} \sum_{b=1}^{B_j} \phi_{bj}(\xvec;T_j, \gamma_j )\phi_{bj}(\xvec^\prime;T_j, \gamma_j ), \label{rpath_covy}
\end{equation}
where $\Theta = \big\lbrace \lbrace T_j, \gamma_j \rbrace_{j= 1}^m, \sigma^2 \big\rbrace$. The conditional variance is given by
\begin{equation}
    \text{Var}\big(Y(\xvec)\mid \Theta\big) = m\tau^2 + \sigma^2.
    \label{rpath_vary}
\end{equation}
\end{theorem_label}
\begin{proof}
Fix the set of trees, bandwidth parameters, and $\sigma^2$ and let $\Theta = \Big\lbrace\lbrace T_j, \gamma_j \rbrace_{j = 1}^m, \sigma^2 \Big\rbrace$. By definition, the conditional covariance between $Y(\xvec)$ and $Y(\xvec^\prime)$ is given by 
\begin{align*}
    \text{Cov}\Big(Y(\xvec),\;Y(\xvec^\prime) \mid \Theta \Big) 
    &= \text{Cov}\Big(\sum_{j=1}^m g(\xvec;T_j,M_j,Z_j) + \epsilon(\xvec), \\&\qquad\qquad
    \sum_{k=1}^m g(\xvec^\prime; T_k,M_k,Z_k) + \epsilon(\xvec^\prime) \mid \Theta\Big) \\[5 pt] 
    &= \sum_{j=1}^m\sum_{k=1}^m\text{Cov}\big( g(\xvec;T_j,M_j,Z_j),\; 
    g(\xvec^\prime; T_k,M_k,Z_k) \mid \Theta\big). 
\end{align*}
where $g(\xvec;T_j, M_j, Z_j) = \sum_{b=1}^{B_j}\mu_{bj}z_{bj}(\xvec)$ and $\xvec \ne \xvec^\prime$.  Due to the conditional independence assumption across the $m$ trees and associated parameters, the covariance simplifies as 
\begin{align}
    \text{Cov}\Big(Y(\xvec),\;Y(\xvec^\prime) \mid \Theta\Big)  
    &= \sum_{j=1}^m \text{Cov}\Big(g(\xvec;T_j,M_j,Z_j),\;
    g(\xvec^\prime; T_j,M_j,Z_j)\mid \Theta\Big), \label{rpath_covm}
\end{align}
since $\text{Cov}\big(g(\xvec;T_j,M_j,Z_j),\;
g(\xvec^\prime; T_k,M_k,Z_k)\mid \Theta\big) = 0$ when $j \ne k$. Finally, we can consider the covariance function within the $\jth$ tree, which simplifies as follows
\begin{align*}
    \text{Cov}\Big(g(\xvec;T_j,M_j,Z_j),\;
    g(\xvec^\prime; T_j,M_j,Z_j)\mid \Theta\Big) 
    &= \text{Cov}\Big(\sum_{b=1}^{B_j}  \mu_{bj}z_{bj}(\xvec),\;
    \sum_{d=1}^{B_j} \mu_{dj}z_{dj}(\xvec^\prime)\mid \Theta\Big) \\[5 pt]
    &= \sum_{b=1}^{B_j} \sum_{d=1}^{B_j} \text{Cov}\Big(\mu_{bj}z_{bj}(\xvec),
    \; \mu_{dj}z_{dj}(\xvec^\prime)\mid \Theta\Big).
\end{align*}
Once again, conditional independence between the terminal node parameters and random path assignments implies $\text{Cov}\big(\mu_{bj}z_{bj}(\xvec),\; \mu_{dj}z_{dj}(\xvec^\prime)\big) = 0$ when $b\ne d$. However, when $b=d$ the covariance within the $\jth$ tree is defined by 
\begin{align*}
    \sum_{b=1}^{B_j}  \text{Cov}\big(\mu_{bj}z_{bj}(\xvec),
    \; \mu_{bj}z_{bj}(\xvec^\prime)\mid \Theta\big) 
    &= \sum_{b=1}^{B_j} E\big[ \mu_{bj}^2z_{bj}(\xvec)z_{bj}(\xvec^\prime) \mid \Theta \big] \nonumber \\ &\qquad- E\big[\mu_{bj}z_{bj}(\xvec)\mid \Theta\big]E\big[ \mu_{bj}z_{bj}(\xvec^\prime)\mid \Theta\big] \nonumber \\
    &= \tau^2 \sum_{b=1}^{B_j} \phi_{bj}(\xvec;T_j, \gamma_j )\;\phi_{bj}(\xvec^\prime;T_j, \gamma_j ), 
\end{align*}
where $E\big[\mu_{bj}z_{bj}(\xvec)\big] = 0$ because each terminal node parameter has mean zero and we assume conditional independence between each $\mu_{bj}$ and $z_{bj}(\xvec)$. Returning to (\ref{rpath_covm}), the covariance between $Y(\xvec)$ and $Y(\xvec^\prime)$ when $\xvec \ne \xvec^\prime$ is then defined as
\begin{equation*} 
\text{Cov}\big(Y(\xvec),\;Y(\xvec^\prime)\mid \Theta \big) = m\tau^2 \sum_{j = 1}^m \frac{1}{m} \sum_{b=1}^{B_j} \phi_{bj}(\xvec;T_j, \gamma_j )\phi_{bj}(\xvec^\prime;T_j, \gamma_j ). 
\end{equation*}
An expression for the conditional variance of $Y(\xvec)$ is also given by
\begin{align}
    \text{Var}\big(Y(\xvec)\mid \Theta\big) &= \text{Var}\big(\sum_{j = 1}^m g(\xvec;T_j,M_j,Z_j) + \epsilon(\xvec) \mid \Theta\big) \nonumber \\
    &= \sum_{j = 1}^m  \text{Var}\big(g(\xvec;T_j,M_j,Z_j)\mid \Theta\big) + \text{Var}\big(\epsilon(\xvec)\mid \Theta\big). \label{rpath_vary_expression}
\end{align}
Due to the conditional independence between parameters across trees, we can once again focus on the variance within each tree separately. The conditional variance for the output of the $\jth$ tree model, is given by
\begin{align*}
    \text{Var}\Big(g(\xvec;T_j,M_j,Z_j)\mid \Theta\Big) &= \text{Var}\Big(\sum_{b=1}^{B_j}\mu_{bj}z_{bj}(\xvec)\mid \Theta\Big) \\[5pt]
    &= E\Big[\Big(\sum_{b=1}^{B_j}\mu_{bj}z_{bj}(\xvec) \Big)^2\mid \Theta\Big] - E\Big[\sum_{b=1}^{B_j}\mu_{bj}z_{bj}(\xvec)\mid \Theta\Big]^2 \\[5pt]
    &= \sum_{b=1}^{B_j}\sum_{d=1}^{B_j} E\Big[\mu_{bj}\mu_{dj}z_{bj}(\xvec)z_{dj}(\xvec)\mid \Theta\Big].
\end{align*}
By assumption, $z_{bj}(\xvec) \in \lbrace 0,1 \rbrace$ and $\sum_{b=1}^{B_j} z_{bj}(\xvec) = 1$. Thus, if $b \ne d$, then $z_{bj}(\xvec)z_{dj}(\xvec) = 0$ with probability 1. The conditional variance then simplifies as 
\begin{align*}
    \text{Var}\big(g(\xvec;T_j,M_j,Z_j)\mid \Theta\big) &= \sum_{b=1}^{B_j} E\big[\mu_{bj}^2 z^2_{bj}(\xvec)\mid \Theta\big] \\[5pt]
    &= \sum_{b=1}^{B_j} E\big[\mu_{bj}^2 z_{bj}(\xvec)\mid \Theta\big] \\[5pt]
    &= \sum_{b=1}^{B_j} \tau^2 \phi_{bj}(\xvec; T_j, \gamma_j ) \\[5pt]
    &= \tau^2,
\end{align*}
where $\sum_{b=1}^{B_j} \phi_{bj}(\xvec; T_j, \gamma_j ) = 1$. Thus, the conditional variance of the tree output is simply the variance of the terminal node parameters $\tau^2$. The variance of the sum-of-tress is then given by $m\tau^2$. These results are the same as in the original BART model. Returning to (\ref{rpath_vary_expression}), the conditional variance of $Y(\xvec)$ is given by $\text{Var}\big(Y(\xvec)\mid \Theta\big) = m\tau^2 + \sigma^2$.

\end{proof}

\subsection{Proof of the Semivariogram Formula} \label{subsect:proof_rpath_svg}

Consider the semivariogram of the sum-of-trees model, which depends on the function $\nu(\xvec,\hvec)$. Due to the constant mean assumption in RPBART, $\nu(\xvec,\hvec)$ is defined by
\begin{align}
    \nu(\xvec, \hvec) &= \frac{1}{2} E^{T,M,Z,\gamma}\Big[\big(Y(\xvec+\hvec)-Y(\xvec)\big)^2\mid \sigma^2\Big]  \nonumber \\[5pt]
    &= \frac{1}{2} E^{T,\gamma}\Big[E^{M,Z}\Big[\big(Y(\xvec+\hvec)-Y(\xvec)\big)^2 \mid \Theta \Big]\Big], \label{rpath_svg_expression}
\end{align} 
where $E^{T,\gamma}$ denotes the expectation with respect to the set of trees and bandwidth parameters and $E^{M,Z}$ denotes the conditional expectation with respect to the set of terminal nodes and random path assignments. Moving forward, we will let $\Theta = \big\lbrace \lbrace T_j, \gamma_j \rbrace_{j= 1}^m, \sigma^2 \big\rbrace$ and treat $\sigma^2$ as a fixed value rather than a random variable.

First consider the inner expectation, which will enable us to understand the effect of the priors we assign to each $M_j$ and $Z_j$ in the model. Due to the constant mean assumption, the inner expectation is equivalently expressed as 
\begin{align}
    E^{M,Z}\Big[\big(Y(\xvec+\hvec)-Y(\xvec)\big)^2 \mid \Theta \Big] &= \text{Var}\big(Y(\xvec+\hvec)\mid \Theta\big) \label{rpath_cond_svg_expression} \\ &\quad\quad + \text{Var}\big(Y(\xvec) \mid \Theta \big) \nonumber\\ &\quad\quad - 2\;\text{Cov}\big(Y(\xvec+\hvec),Y(\xvec) \mid \Theta\big). \nonumber
\end{align}
The inner expectation from  (\ref{rpath_cond_svg_expression}) divided by $2$ will be denoted as $\nu(\xvec,\hvec; \Theta)$. Thus, we can focus on $\nu(\xvec,\hvec; \Theta)$, which involves analytically tractable terms, as shown in Theorem \ref{thm:ccv}:
\begin{equation}
    \nu(\xvec, h;\Theta) = \sigma^2 + m\tau^2 \Big(1 - \frac{1}{m} \sum_{j = 1}^m \sum_{b=1}^{B_j} \phi_{bj}(\xvec+\hvec;T_j, \gamma_j )\phi_{bj}(\xvec;T_j, \gamma_j )\Big) \label{rpath_cond_svg}.
\end{equation}

The function $\nu(\xvec,\hvec)$ is obtained by computing the outer expectation in (\ref{rpath_svg_expression}), which is with respect to the set of $m$ trees and bandwidth parameters. We can then obtain $\nu(\xvec, \hvec)$ by marginalizing over the remaining parameters,
\begin{align}
    \nu(\xvec, \hvec)
    &= \sigma^2 + m\tau^2 E\Big[\Big(1 - \frac{1}{m} \sum_{j = 1}^m \sum_{b=1}^{B_j} \phi_{bj}(\xvec+\hvec;T_j, \gamma_j )\phi_{bj}(\xvec;T_j, \gamma_j )\Big)\Big] \nonumber \\[5 pt]
    &=  \sigma^2 + m\tau^2\Big(1 - \frac{1}{m} \sum_{j = 1}^m E\Big[\sum_{b=1}^{B_j} \phi_{bj}(\xvec+\hvec;T_j, \gamma_j )\phi_{bj}(\xvec;T_j, \gamma_j )\Big]\Big), \label{rpath_uncond_svg}
\end{align}
where the expectation is with respect to $\Theta$ and $\sigma^2$ is treated as a fixed constant.

Since the set of trees and bandwidth parameters
are sets of i.i.d. random quantities, the expectation in (\ref{rpath_uncond_svg}) is the same for each $j = 1,\ldots,m$. Finally, we take $\tau = \big( y_{max} - y_{min}\big)/(2k\sqrt{m})$ which implies $m\tau^2$ simplifies as
\begin{equation*}
    m\tau^2 = \Big(\frac{y_\text{max}-y_\text{min}}{2k}\Big)^2,
\end{equation*}
where $y_\text{max}-y_\text{min}$ is the range of the observed data and $k$ is a tuning parameter that controls the flexibility of the sum-of-trees model. Given these two simplifications, $\nu(\xvec, \hvec)$ for the RPBART is given by  
\begin{align*}
    \nu(\xvec, \hvec)
    &=  \sigma^2 + \Big(\frac{y_\text{max}-y_\text{min}}{2k}\Big)^2\Big(1 - E\Big[\sum_{b=1}^{B_1} \phi_{b1}(\xvec+\hvec;\gamma_1, T_1)\phi_{b1}(\xvec;\gamma_1,T_1)\Big]\Big). 
\end{align*}
\hfill\BlackBox

\subsection{Applications and Interpretations of the Semivariogram}

\begin{figure}[t]
    \centering
    \includegraphics[width = 1\textwidth, height = 1\textwidth]{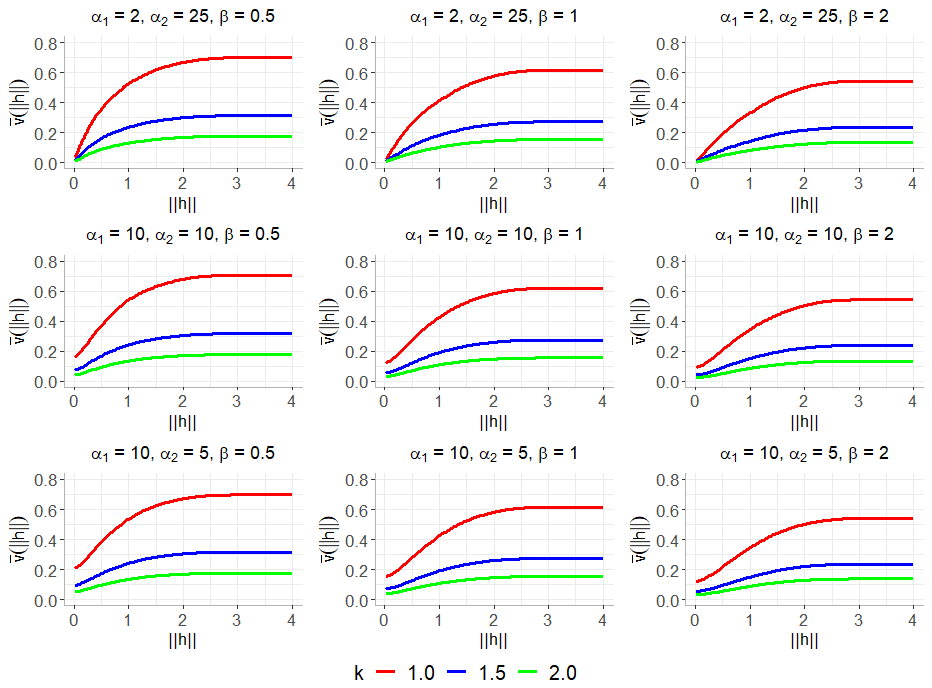}
    \caption{The semivariogram with respect to different hyperparameter settings for the tree, bandwidth, and terminal node parameter priors. Each row corresponds to different settings $\alpha_1$ and $\alpha_2$, which control the bandwidth. Each column corresponds to different values of  $\beta$, which controls the tree size, and $k$. The three lines in each plot represent the semivariogram under different values of $k$. For simplicity, we set $y_{max} = 1$ and $y_{min} = -1$ in the computation of $\tau^2$.}
    \label{fig:rpath_svg_grid}
\end{figure}

A semivariogram is derived for the RPBART model to better understand how the specification of the prior hyperparameters impact the underlying model. For simplicity, consider the two dimensional input space $[-1,1] \times [-1,1]$. Figure \ref{fig:rpath_svg_grid} displays a more detailed example of the impact of the tree prior on the semivariogram. Each row of Figure \ref{fig:rpath_svg_grid} corresponds to a different level of smoothing, where the top row illustrates the lowest amount of smoothing and the bottom row illustrates the highest degree of smoothing. Each column corresponds to a different tree depth, where the leftmost column indicates deeper trees and the rightmost column indicates shallow trees. 

Similar conclusions can be made to those in Section 3.4. Less smoothing and deeper trees result in a more localized fit (row 1 column 1 of Figure \ref{fig:rpath_svg_grid}). In other words, less information is shared across the terminal nodes. 

\begin{itemize}
    \item For each row, the sill of the semivariogram shifts downward as the tree depth decreases. This is because shallower trees result in less terminal nodes. Less terminal nodes increases the probability that a pair of input points separated by a distance $||\hvec||$ are assigned to the same partition of data (and thus naturally more correlated). 
    
    \item The value of $k$ is inversely related to the prior variance for the terminal parameters, $\tau^2$. In the semivariogram, $\tau^2$ simply scales the covariance function. Thus, as $k$ increases, the prior variance decreases and the semivariogram contracts. From a functional perspective, this means less variability in the data is attributed to the underlying mean (modeled as a sum-of-trees) and more variability will be attributed to the random error component.
    
    \item The smoothing parameters generally determine the size of the offset occurring when $||\hvec||$ is near 0. Low levels of smoothing result in a more localized fit. Thus, the correlation between nearby points assigned to adjacent terminal nodes is only non-zero for pairs near the cutpoints. The semivariogram is relatively linear in $||\hvec||$ when $||\hvec||$ is small (top row). This behavior of the semivariogram begins to resemble that of the exponential kernel, which is closely connected to tree kernels \citep{linero2017review}. As more smoothing is introduced, the correlation across points assigned to adjacent terminal nodes increases. In other words, pairs of points can have non-zero correlations even if they are further away from the cutpoints and assigned to different terminal nodes. This causes the offset to shift upwards. In this context, we refer to the ``offset" as the value of the semivariogram as $||\hvec||$ approaches 0. This offset is always positive because of the discontinuity in the RPBART covariance function. Additionally, the higher smoothing case tends to result in a slight curvature in the tail, which is commonly observed with semivariogram that are defined by smooth kernels.

    \item An important theoretical note is that the RPBART model is only continuous in expectation. The discontinuity in the RPBART covariance results from the independence assumption between two random paths, $z_{bj}(\xvec)$ and $z_{bj}(\tilde{\xvec})$. The discontinuity is emphasized when the prior emphasizes a more smooth expectation (rows 2 and 3).

\end{itemize}

\subsubsection{Application}

\begin{figure}[t]
    \centering
    \includegraphics[width = 0.9\textwidth, height = 0.35\textwidth]{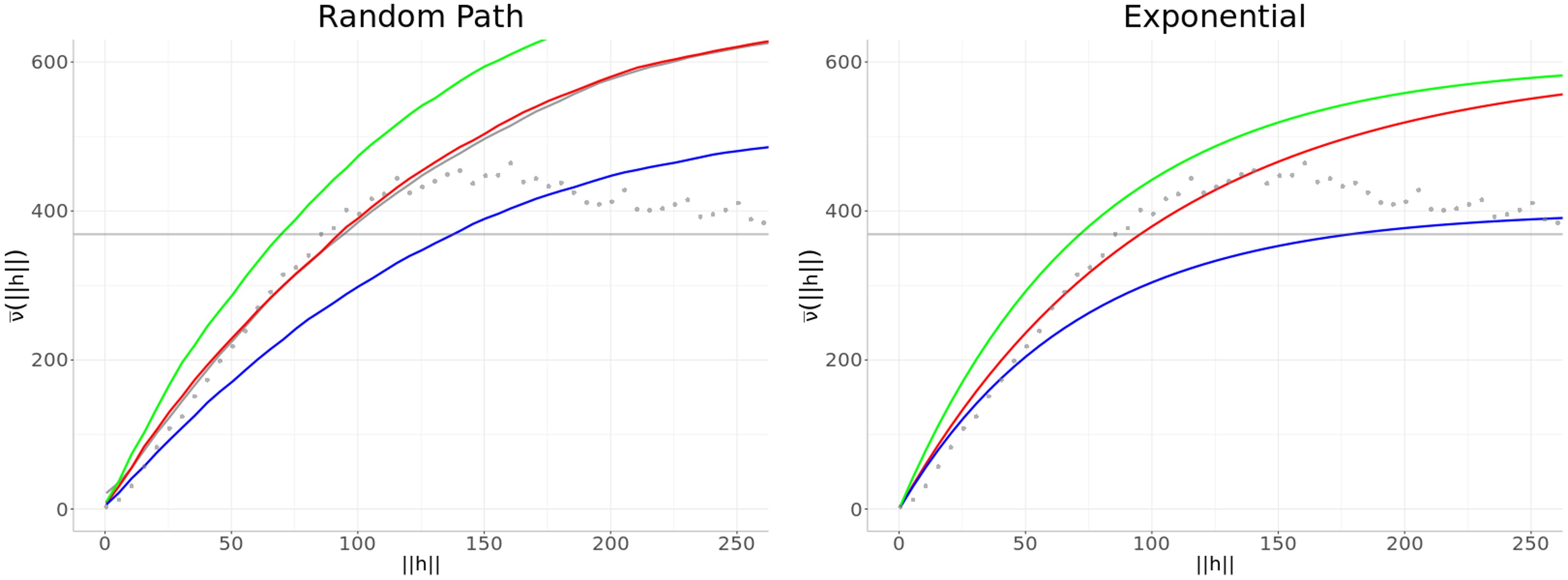}
    \caption{(Left) Four example semivariograms from the random path model using $k = 1.5, \alpha_1 = 2, \alpha_2 = 20, \beta = 1.75$ (dark gray) $k = 1.5, \alpha_1 = 2, \alpha_2 = 65, \beta = 1.75$ (red), $k = 1.7, \alpha_1 = 2, \alpha_2 = 65, \beta = 1.75$ (blue), and $k = 1.5, \alpha_1 = 2, \alpha_2 = 65, \beta = 1.0$ (green) (Right) Examples of the exponential semivariogram with covariance $R(\xvec+\hvec,\xvec;s^2,\ell) = s^2\exp(-||\hvec||/\ell)$ with $(s^2,\ell) = (600,100)$ (red), $(s^2,\ell) = (400,70)$ (blue), and $(s^2,\ell) = (600,75)$ (green). The empirical sill (horizontal gray) and empirical semivariogram (points) are displayed in both plots. }
    \label{fig:miroc-svg}
\end{figure}

We consider emulating the MIROC-ESL2 simulator to demonstrate the semivariogram in practice. This simulator is defined over the domain $[-90^\circ,90^\circ]$ latitude and $[-180^\circ, 180^\circ]$ longitude with points spaced evenly along a $64 \times 128$ grid. For this emulation task, we consider the average monthly temperature output (degrees Celsius) for August 2014. The empirical semivariogram (gray points) along with the model-dependent theoretical semivariograms are shown in Figure \ref{fig:miroc-svg}. The top panel depicts the semivariogram for four possible selections of the RPBART prior. All priors emphasize lower levels of smoothing, which preserves a more localized fit with smoothing very close to a cutpoint. Low smoothing (and thus a more localized fit) is to be expected because this is an emulation task where almost all of the variability in the data should be attributed to the sum-of-trees rather than random error. In such a case, the offset value must be close to 0 for small $||\hvec||$. 

The blue, green, and dark gray curves display the effects of the prior hyperparameters relative to the red curve. The red and dark gray curves compare the impact of $\alpha_2$ on the prior. The gray curve involves relatively more smoothing (due to the smaller $\alpha_2$ value), which results in a small increase in the offset and subtle curvature for $||\hvec||$ near 0. As $||\hvec||$ increases, the two semivariograms are nearly identical. Comparing the red and blue curves allows us to understand the impact of $k$, as larger $k$ decreases the prior variance and thus contracts the semivariogram. Finally, comparing the red and green curves allows us to see the effect of the tree prior, where deeper trees (green) result in less correlation across neighboring points and thus a higher sill. These conclusions agree with the simulated semivariogram in Figure \ref{fig:rpath_svg_grid}.  

The bottom panel of Figure \ref{fig:miroc-svg} displays three example fits for a semivariogram using an exponential kernel defined as $R(\hvec;s^2, \ell) = s^2\exp(-||\hvec||/\ell)$. These different parameter choices were made simply to illustrate the similarities between the final semivariograms under the RPBART model and the exponential kernel.

Overall, we see similar functional forms for the semivariograms under the different models. Both the RPBART and the exponential kernels fit the empirical semivariogram fairly well for $|| \hvec||$ between 0 and 150. The theoretical semivariograms deviate from the empirical semivariogram once reaching the sill. This is not uncommon, as in practice there are less pairs of points separated by large values of $||\hvec||$. Thus, the empirical semivariogram has greater uncertainty for large $||\hvec||$. This example not only illustrates how the RPBART semivariogram can compare to more common semivariograms, but also demonstrates practical challenges with the semivariogram. In practice, one can rarely obtain a perfect fit to the empirical semivariogram. Usually, the practitioner must select elements of the empirical data to fit. In this case, we focus on the fit to the left region of the empirical semivariogram, as we are particularly more concerned with the correlations between points that are nearby, rather than points that are much farther away (i.e. the correlation represented by the sill).

\subsection{Extending the Semivariogram for Model Mixing} \label{subsect:mm_svg}
This section details how the RPBART semivariogram can be extended to model mixing.
\subsubsection{The Semivariogram formula for Model Mixing}

To calibrate the remaining hyperparameters of our RPBART-based mixing model, we employ the prior semivariogram  introduced in Section \ref{subsect:rpath_semivg}.
The semivariogram of Theorem \ref{thm_psvg} can  be extended to the model mixing framework by considering the assumptions associated with each emulator. One cannot directly apply the results conditional on the point estimates of each emulator,  $\hat{f}_1(\xvec),\ldots,\hat{f}_K(\xvec)$, because the semivariogram is designed to assess the spatial variability of the model free of any mean trend \citep{gringarten2001teacher}.  

Rather than condition on the model predictions, we treat each of the individual emulators $f_1(\xvec),\ldots,f_K(\xvec)$ as  unknown quantities. Thus, the semivariogram for $Y(\xvec)$ will assess the modeling choices for the sum-of-trees mixing model along with modeling choices for each $f_l(\xvec)$. For example, we might assume GP emulators,
\begin{align*}
    f_l(\xvec) \mid \psivec_l \; \sim \text{GP} \big(\bar{f}_l,\; R_l(\xvec, \xvec^\prime; \psivec_l) \big),
\end{align*}
where $\psivec_l$ is a vector of parameters such as the scale or length scale and $\bar{f}_l$ is a constant mean.

\begin{theorem_label} \label{thm_psvg_mm}
Assume the random quantities $\lbrace T_j, M_j, Z_j, \gamma_j \rbrace_{j = 1}^m$ are distributed according to Section \ref{subsect:rpath_prior}. Assume each simulator is modeled as a stochastic emulator with mean $\bar{f}_l$ and covariance kernel $R_l({\xvec},\xvec^\prime;\boldsymbol{\psi}_l),\ l=1,\ldots,K.$ Then, the function $\nu(\xvec, \hvec)$ is  
    \begin{align*}
    \nu(\xvec, \hvec)
    &=  \sigma^2 + 
    \Big(\;\frac{1}{4k^2} + \frac{1}{K^2}\Big) 
    \sum_{l = 1}^K\nu_l^{(f)}(\xvec,h;\psivec_l) \; \\ &\quad + \Big(\frac{1}{4k^2}\Big)\Big(1 - \bar{\Phi}(\xvec,\hvec)\Big)\times \sum_{l = 1}^K \Big(R_l(\xvec+\hvec,\xvec;\psivec_l) + \bar{f}_l^2\Big).
    \end{align*}
\end{theorem_label}

The semivariogram, $\overline{\nu}(\lVert \hvec \rVert)$, can be obtained by averaging over $\xvec$, as in (\ref{integral_svg_var}). Rather than  using the empirical semivariogram for $Y(\xvec)$ to select the hyperparameters  for the emulators and the weight functions, we recommend  a modularization approach. Specifically, the hyperparameters associated with each emulator can be selected solely based on evaluations of the corresponding simulator output. The empirical semivariogram  of $Y(\xvec)$ can then be used to select the hyperparameters for the BART weights, plugging-in  the choices of the $\psivec_l$ and $\bar{f}_l$ for $l=1,\ldots,K$. Though this strategy is just used for hyperparameter selection, it aligns well with the two-step estimation procedure common in stacking \citep{breiman_stacking, le2017bayes}, which separates the information used to train the emulators from the observational data used to train the weights. 

\subsubsection{Derivation of the Semivariogram Formula for Model Mixing} \label{subsect:proof_rpath_svg_mm}

Similar to Section \ref{subsect:proof_rpath_svg}, we first consider the function, $\nu(\xvec,\hvec; \Theta)$. In model mixing, this is derived by marginalizing over $M_j$, $Z_j$, and $f_l(\xvec)$ where $j = 1,\ldots,m$ and $l = 1,\ldots,K$. For simplicity, assume each $f_l(\xvec)$ is an independently distributed stochastic process with constant mean $\bar{f}_l$ and covariance function $R_l(\xvec+\hvec, \xvec; \psivec_l)$, with hyperparameter vector $\psivec_l$.

The function $\nu(\xvec,\hvec;\Theta)$ is defined in Theorem \ref{thm:mm_pcsvg}. 
\begin{theorem_label} \label{thm:mm_pcsvg}
Assume the set of emulators, $f_1(\xvec),\ldots,f_K(\xvec)$, are mutually independent random processes with constant means $\bar{f}_l$ and covariance functions $R_l(\xvec+\hvec,\xvec;\psivec_l)$, $l = 1,...,K$. Further assume the emulators are independent of the $K$ weight functions. Then, the function $\nu(\xvec,\hvec;\Theta)$ for the mean-mixing model is given by
    \begin{align*}
        \nu(\xvec,\hvec;\Theta) = \sigma^2 &+ \Big(m\tau^2 + \frac{1}{K^2}\Big) 
        \sum_{l = 1}^K\nu_l^{(f)}(\xvec,h;\psivec_l) \; \\ &\quad + \nu^{(w)}(\xvec,h;\Theta) \sum_{l = 1}^K \Big(R_l(\xvec+\hvec,\xvec;\psivec_l) + \bar{f}_l^2\Big)
    \end{align*}
where $\nu^{(f)}_l(\xvec,h;\psivec_l) = R_l(\xvec,\xvec;\psivec_l) - R_l(\xvec+\hvec,\xvec;\psivec_l)$ is semivariogram component from the $\lth$ emulator, $R(\xvec+\hvec,\xvec+\hvec;\psivec_l) = R(\xvec,\xvec;\psivec_l)$, and $$\nu^{(w)}(\xvec, h;\Theta) = m\tau^2\Big(1 - \frac{1}{m}\sum_{j = 1}^m\sum_{b=1}^{B_j} \phi_{bj}(\xvec+\hvec;T_j, \gamma_j)\phi_{bj}(\xvec;T_j, \gamma_j)\Big)$$ is the semivariogram component from each of the $K$ weight functions.
\end{theorem_label}

\begin{proof}
    
By definition, the function $\nu(\xvec,\hvec; \Theta)$ is defined by 
\begin{align*}
    \nu(\xvec,\hvec; \Theta) &= \frac{1}{2} \text{Var}\Big(Y(\xvec+\hvec)-Y(\xvec) \mid \Theta \Big) \\ 
    &= \frac{1}{2} \bigg(\text{Var}\Big(Y(\xvec+\hvec) \mid \Theta \Big) + \Big(Y(\xvec) \mid \Theta \Big)\bigg) \\
    &\qquad -\text{Cov}\big(Y(\xvec+\hvec)-Y(\xvec)\big.
\end{align*}
The conditional variance simplifies as,    
\begin{align*}
    \text{Var}\Big(Y(\xvec) \mid \Theta \Big) &= \text{Var}\Big(\sum_{l=1}^K w_l(\xvec)f_l(\xvec) + \epsilon(\xvec) \mid \Theta \Big) \\[5pt]
    &=  \sigma^2 + \sum_{l=1}^K \text{Var}\Big( w_l(\xvec)f_l(\xvec) \mid \Theta \Big),
\end{align*}
where the $K$ weights and $K$ functions are all mutually independent, conditional on $\Theta$. Each individual variance component then simplifies as
\begin{align*}
\text{Var}\Big(w_l(\xvec)f_l(\xvec) \mid \Theta \Big) &=
    E\Big[w_l(\xvec)^2f_l(\xvec)^2 \mid \Theta \Big] - E\Big[w_l(\xvec)f_l(\xvec) \mid \Theta \Big]^2 \\[5pt]
    &= E\big[w_l(\xvec)^2\mid \Theta\big] E\big[f_l(\xvec)^2\big] - E\big[w_l(\xvec) \mid \Theta \big]^2E\big[f_l(\xvec)\big]^2 \\[5pt]
    &= \Big(m\tau^2 + 1/K^2\Big)\Big(R_l(\xvec,\xvec;\psivec_l) + \bar{f}_l^2\Big) - \Big(\bar{f}_l/K\Big)^2.
\end{align*}
Similarly, the conditional variance of $w_l(\xvec+\hvec)f_l(\xvec+\hvec)$ is expressed by 
\begin{align*}
\text{Var}\Big(w_l(\xvec+\hvec)f_l(\xvec+\hvec) \mid \Theta \Big) &= \Big(m\tau^2 + 1/K^2\Big) \\ &\times\Big(R_l(\xvec+\hvec,\xvec+\hvec;\psivec_l) + \bar{f}_l^2\Big) \\& - \Big(\bar{f}_l/K\Big)^2,
\end{align*}
for $l=1,\ldots,K$.  \newline\newline The conditional covariance can then be computed as
\begin{align*}
    \text{Cov}\Big(Y(\xvec+\hvec), Y(\xvec) \mid \Theta \Big) &=
    \text{Cov}\Big(\sum_{l=1}^K w_l(\xvec+\hvec)f_l(\xvec+\hvec), \; \sum_{t=1}^K w_t(\xvec)f_t(\xvec) \mid \Theta \Big) \\[5pt]
&=\sum_{l=1}^K\sum_{t=1}^K\text{Cov}\Big(w_l(\xvec+\hvec)f_l(\xvec+\hvec), \;  w_t(\xvec)f_t(\xvec) \mid \Theta \Big).
\end{align*}
Due to conditional independence between the $K$ weight functions and the $K$ emulators, the covariance simplifies as
\begin{align*}
    \text{Cov}\Big(Y(\xvec+\hvec), Y(\xvec) \mid \Theta \Big) &=
    \sum_{l=1}^K\text{Cov}\Big( w_l(\xvec+\hvec)f_l(\xvec+\hvec), \;w_l(\xvec)f_l(\xvec) \mid \Theta \Big) \\[5pt]
&=\sum_{l=1}^K E\Big[w_l(\xvec+\hvec)f_l(\xvec+\hvec)w_l(\xvec)f_l(\xvec) \mid \Theta \Big] \\  &\qquad\quad -E\Big[w_l(\xvec+\hvec)f_l(\xvec+\hvec) \mid \Theta \Big]E\Big[w_l(\xvec)f_l(\xvec) \mid \Theta \Big] \\[5pt]
&=\sum_{l=1}^K E\Big[w_l(\xvec+\hvec)w_l(\xvec)\mid\Theta\Big] E\Big[f_l(\xvec+\hvec)f_l(\xvec)\Big] - \Big(\bar{f}_l/K\Big)^2 \\[5pt]
&= \sum_{l=1}^K \big(m\tau^2\bar{\phi}(\xvec,h;\Theta) + 1/K^2\big)\Big(R_l(\xvec+\hvec,\xvec;\psivec_l) + \bar{f}_l^2\Big) \\
&\qquad \quad- \sum_{l=1}^K \Big(\bar{f}_l/K\Big)^2.
\end{align*}
where $\bar{\phi}(\xvec,h;\Theta) = \frac{1}{m}\sum_{j = 1}^m \sum_{b = 1}^{B_j} \phi_{bj}(\xvec+\hvec;T_j,\gamma_j)\phi_{bj}(\xvec;T_j,\gamma_j)$. Using the conditional variances and covariance, $\nu(\xvec,\hvec; \Theta)$ can be expressed as
\begin{align*}
    \nu(\xvec,\hvec; \Theta) &= \frac{1}{2}\bigg(\sigma^2 + \sum_{l=1}^K \Big(R_l(\xvec+\hvec,\xvec+\hvec;\psivec_l) + \bar{f}_l^2\Big)\Big(m\tau^2 + 1/K^2\Big) - \Big(\bar{f}_l/K\Big)^2 \bigg) \\[5pt]
    &\quad + \frac{1}{2}\bigg(\sigma^2 + \sum_{l=1}^K \Big(R_l(\xvec,\xvec;\psivec_l) + \bar{f}_l^2\Big)\Big(m\tau^2 + 1/K^2\Big) - \Big(\bar{f}_l/K\Big)^2 \bigg) \\[5pt]
    &\quad - \Bigg(\sum_{l=1}^K \big(m\tau^2\bar{\phi}(\xvec,h;\Theta)  + 1/K^2\big)\Big(R_l(\xvec+\hvec,\xvec;\psivec_l) + \bar{f}_l^2\Big) - \Big(\bar{f}_l/K\Big)^2\Bigg) \\[8pt]
\end{align*}
This expression can then be simplified as follows:

\begin{align*}
    \nu(\xvec,\hvec; \Theta) &= \sigma^2 + \frac{1}{2}\bigg(\sum_{l=1}^K \Big(R_l(\xvec+\hvec,\xvec+\hvec;\psivec_l)\Big)\Big(m\tau^2 + 1/K^2\Big) + m\tau^2\bar{f}^2_l \bigg) \\[5pt]
    &\quad + \frac{1}{2}\bigg(\sum_{l=1}^K \Big(R_l(\xvec,\xvec;\psivec_l)\Big)\Big(m\tau^2 + 1/K^2\Big) + m\tau^2\bar{f}^2_l \bigg) \\[5pt]
    &\quad - \Bigg(\sum_{l=1}^K \big(m\tau^2\bar{\phi}(\xvec,h;\Theta)  + 1/K^2\big)R_l(\xvec+\hvec,\xvec;\psivec_l) + m\tau^2\bar{f}_l^2\bar{\phi}(\xvec,h;\Theta)\Bigg) \\[8pt]    
    &= \sigma^2 + \frac{1}{2}\Big(m\tau^2 + 1/K^2\Big)\bigg(\sum_{l=1}^K \Big(R_l(\xvec+\hvec,\xvec+\hvec;\psivec_l) + R_l(\xvec,\xvec;\psivec_l)\bigg) \\[5pt] 
    &\qquad \quad + m\tau^2\big(1 - \bar{\phi}(\xvec,h;\Theta)\big)\sum_{l = 1}^K \bar{f}_l^2  \\
    &\qquad \quad - \big(m\tau^2\bar{\phi}(\xvec,h;\Theta)  + 1/K^2\big)\sum_{l = 1}^K  R_l(\xvec+\hvec,\xvec;\psivec_l) \\
    &\qquad \quad \pm \Big(m\tau^2 + 1/K^2\Big) \sum_{l=1}^K R_l(\xvec+\hvec, \xvec; \psivec_l),
\end{align*}
where we can add and subtract $\Big(m\tau^2 + 1/K^2\Big) \sum_{l=1}^K R_l(\xvec+\hvec, \xvec;\psivec_l)$ to further simplify the expression. Now define the following functions for each emulator and each weight as
\begin{align}
    \nu^{(f)}_l(\xvec,\hvec;\psivec_l) &= \frac{1}{2}\Big(R_l(\xvec,\xvec;\psivec_l) + R_l(\xvec+\hvec,\xvec+\hvec;\psivec_l) - 2R_l(\xvec+\hvec,\xvec;\psivec_l)\Big) \label{append:pwsvgf} \\
    \nu^{(w)}(\xvec, \hvec;\Theta) 
    &= m\tau^2\Big(1 - \bar{\phi}(\xvec,h;\Theta)\Big) \label{append:pwsvgw}
\end{align}

In typical cases, $R(\xvec+\hvec,\xvec+\hvec;\psivec_l) = R(\xvec,\xvec;\psivec_l)$ and thus $\nu^{(f)}_l(\xvec,h;\psivec_l)$ simplifies further. The terms in the expression can then be rearranged as follows
\begin{align*}
    \nu(\xvec,\hvec; \Theta) &= \sigma^2 + \Big(m\tau^2 + 1/K^2\Big)\sum_{l=1}^K \nu^{(f)}_l(\xvec,\hvec;\psivec_l) \\
    &\qquad \quad + \nu^{(w)}(\xvec, \hvec;\Theta) \bigg(\sum_{l = 1}^K  R_l(\xvec+\hvec,\xvec;\psivec_l) + \bar{f}_l^2\bigg).
\end{align*}

\end{proof}

Theorem \ref{thm:mm_pcsvg} shows the function $\nu(\xvec,\hvec;\Theta)$ combines information from the model set and the weight functions. More specifically, we see the expression combines the semivariogram functions from the $K+1$ different components. Using Theorem \ref{thm:mm_pcsvg}, we can compute the $\nu(\xvec,\hvec)$ in Definition \ref{thm_psvg} by marginalizing over the set of $m$ trees and bandwidth parameters. The resulting function is given by
\begin{align*}
    \nu(\xvec, \hvec)
    &=  \sigma^2 + 
    \Big(\;\frac{1}{4k^2} + \frac{1}{K^2}\Big) 
    \sum_{l = 1}^K\nu_l^{(f)}(\xvec,h;\psivec_l) \; \\ &\quad + \Big(\frac{1}{2k}\Big)^2\Big(1 - E\Big[\sum_{b=1}^{B_1} \phi_{b1}(\xvec+\hvec;T_1,\gamma_1)\phi_{b1}(\xvec;T_1,\gamma_1)\Big]\Big) \\ &\qquad\quad \times \sum_{l = 1}^K \Big(R_l(\xvec+\hvec,\xvec;\psivec_l) + \bar{f}_l^2\Big).
    \end{align*}
\hfill\BlackBox

We note, the proof of Theorem \ref{thm_psvg_mm} follows directly from Theorem \ref{thm:mm_pcsvg} with the additional step of taking the expectation with respect to $\Theta$.

\subsection{Algorithm}
The sampling algorithm for the random path model is given by Algorithm \ref{alg:rpath_bart}.

\begin{algorithm}
\caption{The sampling algorithm for the random path model.}\label{alg:rpath_bart}
\textbf{Data:} $(\xvec_1,y_1),\ldots,(\xvec_n,y_n)$ \\
\textbf{Result:} Approximate posterior samples drawn from $\pi\Big(\big\{(T_j,M_j,Z_j,\gamma_j)\big\}_{j=1}^m, \sigma^2 \mid \yvec \Big)$
\begin{algorithmic}
\For{$N_{mcmc} \; iterations$} 
    \For{$j=1,\ldots,m$}
        \State (1) With probability $p_{\text{birth}}$, draw $T_j$ and partially update $Z_j$ using RJMCMC. \\ \hspace{18mm} Otherwise update via a death step.
        \State (2) Update each splitting rule $(v_{dj},c_{dj})$ individually for $d = 1,\ldots,B_j - 1$, \\ 
        \hspace{18mm} where $T_j$ has $B_j-1$ internal nodes.
        \State (3) Draw $Z_j \mid T_j,\gamma_j,\rvec_j, \sigma^2$
        \State (4) Draw $M_j \mid T_j, Z_j, \gamma_j, \rvec_j, \sigma^2$
        \State (5) Draw $\gamma_j \mid T_j, M_j, Z_j,\rvec_j, \sigma^2$
    \EndFor
    \State (6) Draw $\sigma^2 \mid \yvec, \big\{(T_j,M_j),Z_j,\gamma_j\big\}_{j=1}^m$
    \EndFor
\end{algorithmic}
\end{algorithm}

Algorithm \ref{alg:rpath_bart} involves conditional sampling steps for each of the components of the RPBART model. Two main differences exist between the RPBART algorithm and previous BART sampling algorithms \citep{pratola2016efficient, bart_2010}. The first difference is in updating the tree structure, as the dimension of $Z_j$ will also change as the dimension of $T_j$ changes. Each tree, $T_j$, can be updated via a birth step, which adds one new terminal node, or a death step, which removes a pair of terminal nodes. Both of these moves will have analogous effects on the dimension the $n$ vectors within $Z_j$; a birth step will increase the dimension by one, while a death step will decrease the dimension by one. A simple reversible jump MCMC (RJMCMC) is used to navigate the dimension change within $Z_j$ when jointly updating $$(T_j,Z_j) \rightarrow (T^\prime_j, Z^\prime_j)$$ where $T^\prime_j$ is the proposed tree and $Z^\prime_j$ is the set of proposed random path vectors.

\subsubsection{Birth Step:}

Without loss of generality, consider a birth proposal that splits the first terminal node $\eta^{(t)}_{1j}$ using the rule $x_v < c$. The proposed move adds one more dimension to each random path vector. Thus, we define the proposed set $Z_j^\prime$ by 
\begin{align*}
    Z_j^\prime &= \Big\lbrace \zvec_{j}^\prime(\xvec_i) \Big\rbrace_{i = 1}^n \\
    \zvec^\prime(\xvec_i) &= \Big(z^{(l)}_{1j}(\xvec_i),\;z^{(r)}_{1j}(\xvec_i),\; z^\prime_{2j}(\xvec_i),\ldots,z^\prime_{B_jj}(\xvec_i) \Big),
\end{align*}
where $z^{(l)}_{1j}(\xvec_i)$ and $z^{(r)}_{1j}(\xvec_i)$ are the new random path indicators for the proposed left and right child nodes. The other components $z^\prime_{bj}(\xvec_i), \; b = 2,\ldots,B_j,$ are the random path indicators corresponding to the terminal nodes which are unaffected by the birth proposal.

Since the tree proposal only affects the first terminal node, we assume the proposal to the random path vectors only affects the first component of the vector. As a result, we set $z^\prime_{bj}(\xvec_i) = z_{bj}(\xvec_i)$ for $i=1,\ldots,n$ and $b=2,\ldots,B_j$ and only consider an update for the first component $z_{1j}(\xvec_i)$. The proposed move can be simplified as 
\begin{equation}
    \Big(T_j,\; \lbrace z_{1j}(\xvec_i)\rbrace_{i=1}^n \Big) \; \rightarrow \; \Big(T_j^\prime,\;\lbrace z^{(l)}_{1j}(\xvec_i), \; z^{(r)}_{1j}(\xvec_i)\rbrace_{i=1}^n \Big), \label{zbirth}
\end{equation}
where $z_{1j}(\xvec_i)$ is the random path indicator for $\eta^{(t)}_{1j}$. 

To facilitate the RJMCMC required by (\ref{zbirth}), we introduce a set of $n$ independent augmented random variables, $u_{1j},\ldots,u_{nj}$, each of which can be interpreted as the event that the $\ith$ observation is randomly mapped to the new right child node. A simple strategy is to set $u_{ij} = 1$ if $x_{iv} \geq c_v$ and $u_{ij} = 0$ otherwise. This strategy imposes a deterministic split to assess the impact of adding a new terminal node to the tree without any additional stochasticity of the random path assignments. A more generalized strategy is further discussed in \citep{yannotty2024bayesian}.  

The augmented random variables are then used to match the dimensions in the proposal as shown in (\ref{rjmcmc_equal_cond}),
\begin{align}
    z^{(r)}_{1j}(\xvec_i) &= u_{ij}\;z_{1j}(\xvec_i) \label{rjmcmc_right_cond} \\
    z^{(l)}_{1j}(\xvec_i) &= (1-u_{ij})\;z_{1j}(\xvec_i) \label{rjmcmc_left_cond} \\
    z^\prime_{bj}(\xvec_i) &= z_{bj}(\xvec_i), \quad b = 2,\ldots,B_j \label{rjmcmc_equal_cond}.
\end{align}

\subsubsection{Death Step:}

The death proposal can be viewed as the reversal of the birth proposal. Without loss of generality, assume $T^\prime_j$ has $B_j + 1$ terminal nodes denoted by $\eta^{(l)}_{1j}$, $\eta^{(r)}_{1j}$, and $\eta^{(t)}_{bj}$ for $b=2,\ldots,B_j$. Now suppose the death move reconstructs $T_j$ by replacing $\eta^{(l)}_{1j}$ and $\eta^{(r)}_{1j}$ with $\eta^{(t)}_{1j}$. This proposal is summarized by 
\begin{equation}
     \Big(T_j^\prime, \; \big\lbrace z^{(l)}_{1j}(\xvec_i), \; z^{(r)}_{1j}(\xvec_i) \big\rbrace_{i=1}^n \Big) \; \rightarrow \; \Big(T_j, \; \big\lbrace z_{1j}(\xvec_i) \big\rbrace_{i=1}^n \Big). \label{zdeath}
\end{equation}
Once more, we wish to fix the remainder of the tree in $T_j^\prime$ and focus solely on the death proposal at the specified terminal nodes. In such a case, no augmentation is needed and the deterministic map is given by 
\begin{align*}
    z_{1j}(\xvec_i) &= z^{(r)}_{1j}(\xvec_i) + z^{(l)}_{1j}(\xvec_i) \\
    z_{bj}(\xvec_i) &= z^\prime_{bj}(\xvec_i), \quad  b = 2,\ldots,B_j
\end{align*}
for $i = 1,\ldots,n$. This implies that only the observations that are mapped to either the left or right children under consideration will affect the evaluation of the death proposal.

\subsubsection{Updating $Z_j$}

Given the update to the tree in Step 1, we further update the random path assignments using a Metropolis-Hastings algorithm. In such a case, we simply use the prior distribution as a proposal distribution. In some cases, a modularized approach can be used to obtain more efficient updating of $Z_j$ \citep{yannotty2024bayesian, liu2009modularization}. The modularized approach ignores the information in the backfitting residuals (which is used to update all other elements of the model) and simply updates $Z_j$ by only considering the current information in $Z_j$, the selected proposal distribution, the updated values of the bandwidth parameters, and the updated tree structure.    

\subsection{Supporting Examples} \label{subsect:examples}
This section provides additional comparisons between RPBART and other competing methods in both regression and model mixing tasks.
\subsubsection{RPBART Regression Example}
Consider the true data generating process 
\begin{align*}
    Y(\xvec) &\sim N\big( f_\dagger(\xvec), \sigma^2 \big) \\
    f_\dagger(\xvec) &= \sin(x_1) + \cos(x_2)
\end{align*} 
where $\xvec \in [-\pi,\pi] \times [-\pi,\pi]$. Assume $n = 100$ different observations are generated from the true underlying process with $\sigma = 0.1$. The inputs $\xvec_1,\ldots,\xvec_{100}$ are also randomly generated about the two-dimensional input space. We train the random path model using three different settings of the bandwidth prior to demonstrate the smoothing effect of the model. Each of the three models are trained using an ensemble of 20 trees and a value of $k = 1$.
\begin{figure}[H]
    \centering
    \includegraphics[width = 0.95\textwidth, height = 1.1\textwidth]{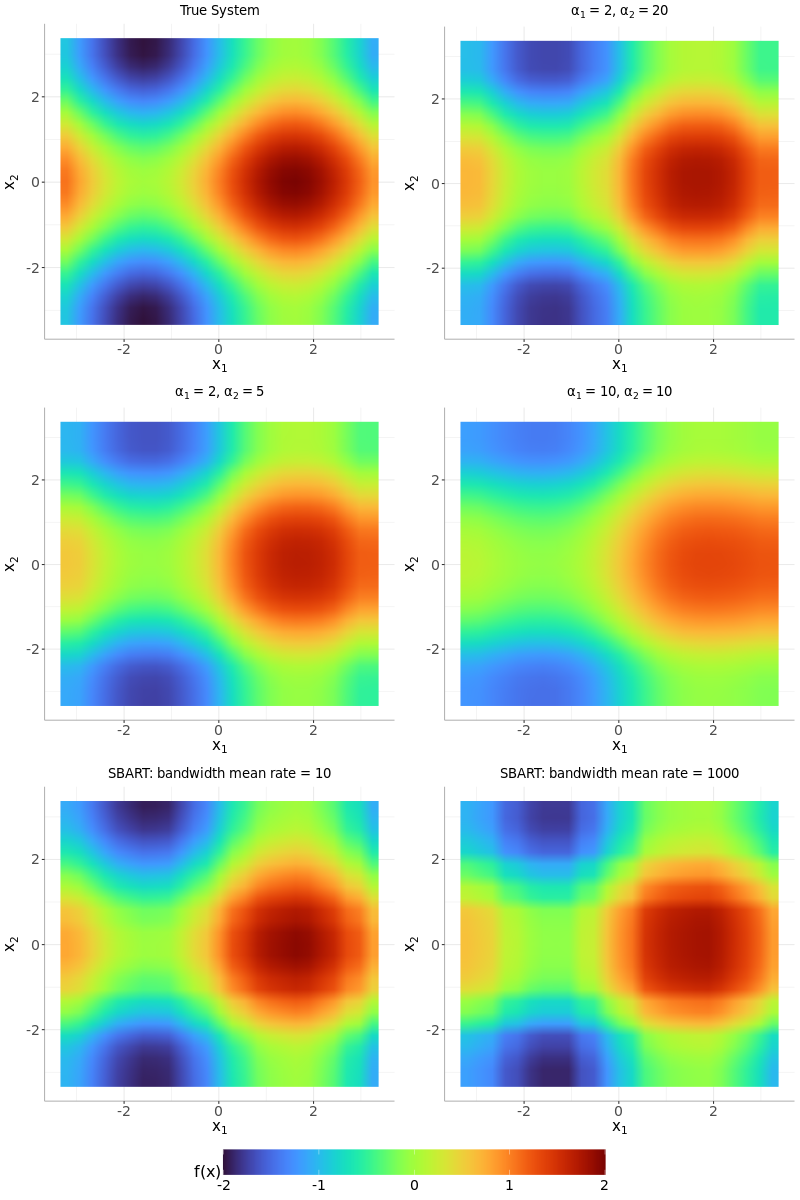}
    \caption{The true system (top left) compared to the the posterior mean predictions from RPBART with different settings of the bandwidth prior as indicated by the values of $\alpha_1$ and $\alpha_2$ (top right and middle row). Two mean predictions from SBART are included with a bandwidth mean rates of 10 and 1000, respectively (bottom row).}
\label{fig:supplement_2dtaylor}
\end{figure}

Figure \ref{fig:supplement_2dtaylor} displays the mean predictions from the RPBART model along with their associated hyperparameters of $\alpha_1$ and $\alpha_2$. Based on Figure \ref{fig:supplement_2dtaylor}, we see the model with less smoothing (i.e. $\alpha_1 = 2$ and $\alpha_2 = 20$) provides the most accurate predictions across the entire domain. Essentially, this model preserves the localization induced by the tree models but is able to smooth over the hard boundaries. With $\alpha_1 = 2$ and $\alpha_2 = 5$, we see the a similar smooth mean surface, however the prediction is slightly less accurate along the boundaries (particularly when $x_1$ is near $-\pi$). We note BART typically struggles along the boundaries when no or minimal data is present, so this occurrence is to be expected. Similar observations can be made as more smoothness is introduced into the model as shown in the middle right panel of Figure \ref{fig:supplement_2dtaylor}, where $\alpha_1 = 10$ and $\alpha_2 = 10$. 

These results are to be expected as higher levels of smoothing will naturally result in a less localized prediction. Despite this, RPBART is able to consistently identify the main features of the underlying process, regardless the level of smoothing. In cases with more smoothing, the mean function will identify the general pattern of the data and attribute some of the granularity in the process to the observational error.    

For comparison, two mean predictions from the SBART model are also displayed in the bottom row of Figure \ref{fig:supplement_2dtaylor}. In SBART, the bandwidth parameter is assigned an exponential prior with rate parameterization. Thus, higher values of the bandwidth lead to less smoothing. In terms of predictive performance, we observe SBART is slightly more accurate in predicting the peaks (dark red) and valleys (dark blue) of the function. Despite this, the mean prediction from SBART still exhibits a block-like structure, which is commonly observed in BART due to the discontinuity of the trees. This most notably occurs near the peak centered around $(2,0)$. Finally, the two SBART predictions are fairly similar, which suggests the model is less sensitive to the prior specification compared to RPBART. Meanwhile, RPBART produces a smoother mean prediction across the domain. The degree of smoothing can be easily controlled through the prior distribution.      

\subsubsection{Comparing BART-BMM and RPBART-BMM}

\begin{figure}[H]
    \centering
    \includegraphics[width = 0.45\textwidth, height = 0.475\textwidth]{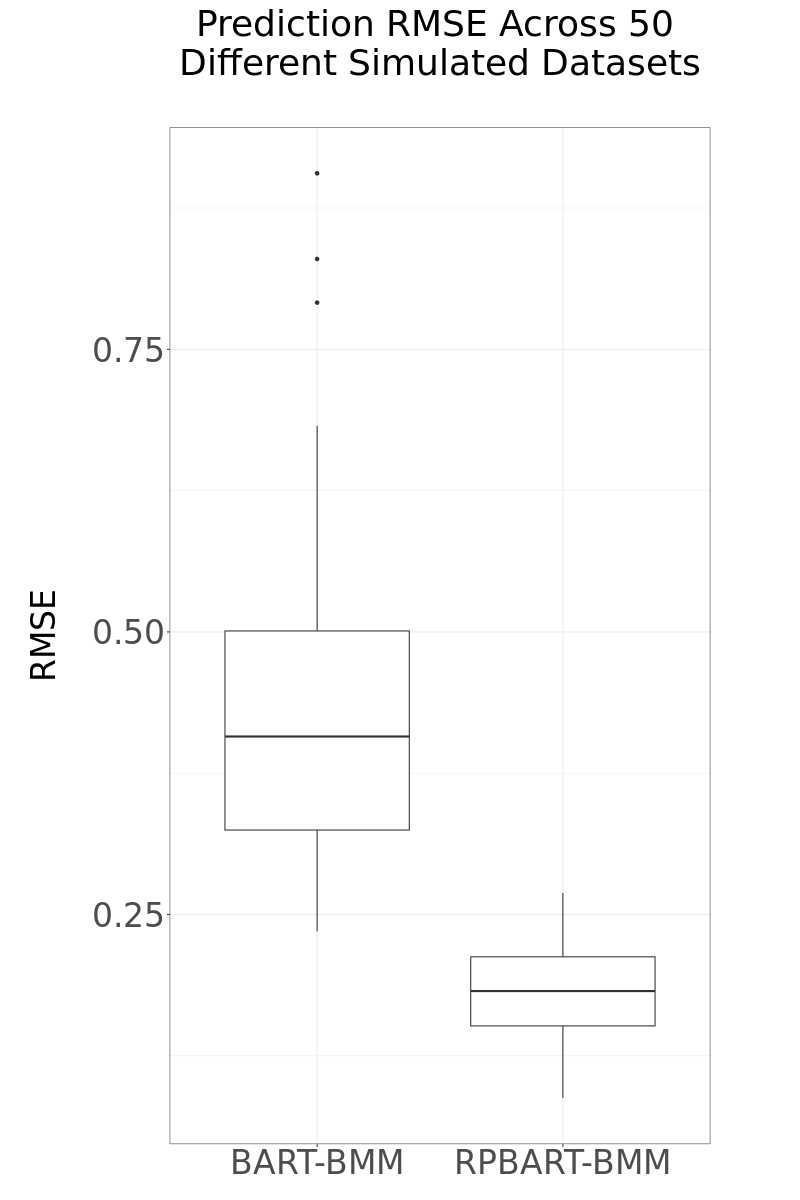}
    \caption{The distribution of the prediction RMSE for BART-BMM and RPBART-BMM across 50 different simulated datasets.}
    \label{fig:rmse_boxplot}
\end{figure}

In this section, we revisit the 2-dimensional Taylor series mixing example by comparing the RPBART-BMM and BART-BMM results across 50 different datasets simulated from the same underlying system. Figure \ref{fig:rmse_boxplot} compares the distribution of the prediction RMSE of BART-BMM versus the prediction RMSE of RPBART-BMM. Clearly, RPBART-BMM outperforms BART-BMM across the simulation study. Ultimately, RPBART-BMM is less prone to overfitting in the region where both simulators are inaccurate and minimal data is available.  

\subsubsection{Climate Results} \label{gcm_table}

This section formally compares the results of the three mixing methods (RPBART-BMM, NNS, and FWLS) and a simple average of the 8 GCMs to those of the individual GCMs. Table \ref{table:cmip_results} illustrates that each mixing approach substantially improves prediction accuracy compared to an individual GCM.

\begin{table}[ht]
\centering
\begin{singlespace} 
\begin{tabular}{ |p{3cm}||c|c|c|  }
 \hline
 Model&April 2014& August 2014 &December 2014\\
 \hline
 RPBART-BMM   & {\bf 0.827}    & {\bf 0.864} &   {\bf 0.882}\\
 NNS   & 0.993    &1.032&  1.088 \\
 FWLS   & 2.393    &2.345 & 2.142\\
 Average & 2.826 & 3.089 & 2.960\\
\hdashline
 Access   & 3.675    &3.813&  4.802\\
 BCC&   4.229  & 4.320   &4.340\\
 MIROC &5.529 & 5.590&  6.979\\
 CMCC    &4.454 & 4.272&  6.369\\
 CESM2&   3.230  & 3.681& 3.419\\
 CNRM& 3.289  & 3.053   &3.580\\
 Can-ESM5& 3.665  & 3.977&3.759\\
 KIOST& 3.763  & 4.645& 3.639\\
 \hline
\end{tabular} 
 \caption{The prediction root mean square error for each mixing approach (RPBART-BMM, NNS, and FWLS), an average of the 8 GCMs, and the individual GCMs (remaining rows) stratified by month. Each RMSE is computed by evaluating the associated model-based prediction over a dense grid of 259,200 inputs and comparing to the observed ERA5 data.  The smallest RMSEs for each month are denoted in bold.}
 \label{table:cmip_results}
\end{singlespace}
\end{table}

\end{document}